\documentclass[11pt]{article}

\usepackage{epsfig}
\begin{document}

\baselineskip=14pt plus 0.2pt minus 0.2pt
\lineskip=14pt plus 0.2pt minus 0.2pt

\newcommand{\hlf}{\mbox{$1\over2$}}
\newcommand{\lfrac}[2]{\mbox{${#1}\over{#2}$}}
\newcommand{\scsz}[1]{\mbox{\scriptsize ${#1}$}}
\newcommand{\tsz}[1]{\mbox{\tiny ${#1}$}}


\begin{center}
\Large{{\bf The Schr\"odinger System} 
$\mathbf{H=-\lfrac{1}{2}\left({{t_o}\over{t}}\right)^a\partial_{xx}+
\lfrac{1}{2}\omega^2\left({{t}\over{t_o}}\right)^bx^2} $}
\vspace{0.25in}

\normalsize

Michael Martin Nieto\footnote{Email:  mmn@lanl.gov}

\vskip 20pt

{\it Theoretical Division (MS-B285), Los Alamos National Laboratory,
University of California, 
Los Alamos, New Mexico 87545, U.S.A.}

\vskip 20pt

 and 

\vskip 20pt

D. Rodney Truax\footnote{Email:  truax@ucalgary.ca}

\vskip 20pt

{\it Department of Chemistry,
 University of Calgary,
Calgary, Alberta T2N 1N4, Canada}

\vskip 20pt
\today

\vspace{0.3in}
 
\end{center}

\baselineskip=.33in

We attack the specific time-dependent Hamiltonian problem 
${H}=-\lfrac{1}{2}\left({{t_o}\over{t}}\right)^a\partial_{xx}+
\lfrac{1}{2}\omega^2\left({{t}\over{t_o}}\right)^bx^2$.
This corresponds to a time-dependent mass ($TM$) Schr\"odinger
equation.  We give the specific transformations to a
different time-dependent 
quadratic Schr\"odinger equation ($TQ$) and to a different time-dependent 
oscillator ($TO$) equation.  For each Schr\"odinger system, 
we give the Lie algebra of space-time symmetries, 
the number states, 
the squeezed-state  $\langle x\rangle$ and   
$\langle p\rangle$ (with their classical motion), 
$(\Delta x)^2$, $(\Delta p)^2$, and the uncertainty product.

\vspace{1.25in}

\newpage

\normalsize

\baselineskip=.33in


\section{INTRODUCTION}

In recent work  \cite{paperI,paperII} we discussed  
general time-dependent quadratic ($TQ$) Schr\"odinger equations 
\begin{eqnarray}
S_1\Phi(x,t) & = & \{-\left[1+k(t)\right]P^2+2T+h(t)D+
                       g(t)P    \nonumber\\
               &   & \hspace{1.0cm}-2h^{(2)}(t)X^2-2h^{(1)}(t)X
                    -2h^{(0)}(t)I\}\Phi(x,t) = 0. \label{e:pre12}
\end{eqnarray}
It was shown how to solve them by  i) first performing a
unitary transformation to a time-dependent mass ($TM$) equation, 
and then ii) making a change of time variable to yield a 
time-dependent oscillator
($TO$) equation which in principle can be solved.  One can then work  
backwards to find the $TM$ and $TQ$ solutions.

Elsewhere \cite{TMI}, 
we went into detail on how to solve a specific subclass 
of cases, $TM$ equations with only time-dependent $P^2$ and $X^2$ 
terms. [We will refer to equations from this paper as, e.g., Eq. 
(\cite{TMI}-22).] 
These Hamiltonians are parametrized as  
\begin{equation}
\hat{H}_2=\lfrac{1}{2}e^{-2\nu(t)}P^2+h^{(2)}(t)e^{2\nu(t)}X^2.
\label{genH2}
\end{equation} 
In this paper we examine the $TM$ system 
\begin{eqnarray}
\hat{H}_2&=&\lfrac{1}{2}\left({{t_o}\over{t}}\right)^a P^2
+\lfrac{1}{2}\omega^2\left({{t}\over{t_o}}\right)^bX^2,   \label{ex2H}
\end{eqnarray}
where $a$ and $b$ are real numbers. (We have also investigate another 
$TM$ system \cite{TMII}

Going directly to what is this problem's $TQ$  Schr\"odinger equation, 
\begin{eqnarray}
TQ: \hspace{2cm}
S_1\Phi(x,t)&=&\left\{-P^2+2T-\frac{a}{t}D-\omega^2\left(\frac{t}{t_o}
\right)^{b-a}X^2\right\}\Phi(x,t)=0,   \label{ex2S1}  \\
H_1&=&\frac{1}{2}P^2 +\frac{1}{2}\frac{a}{t}D 
            +\frac{1}{2}\omega^2\left(\frac{t}{t_o}\right)^{b-a}X^2,   
\label{ex2S1H}
\end{eqnarray}
where the operators $P^2$, $T=i\partial_t$, $D=(XP+PX)/2$, 
and $X^2$ are defined in equations  
(\cite{TMI}-1).  Note that $t_o\ne 0$ and furthermore $t$ and $t_o$ 
must be of the same sign.  For this paper, we assume that $t_o>0$.
This $TQ$ equation is related to the time-dependent 
$TM$ Schr\"odinger equation by the unitary mapping \cite{paperI}
\begin{equation}
R(0,\nu,0)=\exp\left[i\nu D\right],~~~~~~~~~
\nu=\frac{a}{2}\ln{\left(\frac{t}{t_o}\right)}.\label{ex2nu}
\end{equation}
[See Eq. (\cite{paperI}-78,88).] This yields the $TM$ equation 
\begin{eqnarray}
TM: \hspace{2cm}
\hat{S}_2\hat{\Theta}(x,t)&=&\left\{-\left({{t_o}\over{t}}\right)^aP^2
+2T-\omega^2\left({{t}\over{t_o}}\right)^bX^2\right\}\hat{\Theta}
(x,t)=0,\label{ex2S2}
\end{eqnarray}
where $a$ and $b$ are real numbers.
This is the defining  $TM$ equation of our system. 
It has been studied by Kim \cite{spk1} and also in \cite{paperI}. 
The $TM$ equation (\ref{ex2S2}) is transformed into a 
$TO$ equation  
\begin{eqnarray}
TO: \hspace{2cm}
S_3\Psi(x,t')&=&\left\{-P^2+2T-2g^{(2)}(t')X^2\right\}\Psi(x,t')=0,
\label{ex2S3} \\
H_3&=&\lfrac{1}{2}P^2 + g^{(2)}(t')X^2,
\label{ex2HH}
\end{eqnarray} 
by a change in time variable.  When $a=0$,  we trivially have $t=t'$, 
a case we now ignore.    When $a=1$ {\bf (Case 1)}, 
the time transformation is [see Eq. (\cite{paperI}-95)]
\begin{equation}
t'-t_o'=t_o\ln{\left(\frac{t}{t_o}\right)},   
\hspace{1in}
g^{(2)}(t')=\lfrac{1}{2}\omega^2\exp\left[\frac{1+b}{t_o}(t'-t_o')
\right],
\label{ex2ttfna=1}
\end{equation}
with $t'-t_o'\in [0,\infty)$.  When $a\ne 1$ {\bf (Case 2)}, 
the time transformation is [see Eq. (\cite{paperI}-96)]
\begin{equation}
t'-t_o'=\frac{t_o}{1-a}\left[\left(\frac{t}{t_o}\right)^{1-a}-1\right],
\hspace{1cm}
g^{(2)}(t')=\lfrac{1}{2}\omega^2\left[1+
\left(\frac{1-a}{t_o}\right)(t'-t_o')\right]^{\frac{a+b}{1-a}},
\label{ex2pt12}
\end{equation} 
with $t'-t_o'\in [0,t_o/(1-a))$ if $a\in (1,\infty)$ and 
$t'-t_o'\in [0,\infty)$ if $a\in (-\infty,0)\cup (0,1)$.

In Section 2, we show how to derive the time-dependent $TO$ functions, 
these forming the basic solutions to the problem.     
We describe how the two Cases are subdivided into 
different (sub)classes of solutions, depending on the 
different values of the parameters $a$ and $b$.  (The same is done 
later for the $TM$ and $TQ$ solutions. 
In Section 3, we describe the $TO$ solutions, listing all 
the function solutions for the different (sub)classes in an 
Appendix.  (Two examples are given.)  
The same is done for the $TM$ and $TQ$ functions.
With these functions we  compute, in Section 4, 
the time-dependent Lie symmetries that form the bases of 
the oscillator algebras.    

As shown in Ref. \cite{TMI} (and also demonstrated 
in Ref. \cite{TMII}), the explicit calculations of the 
number, coherent, and  squeezed states 
is a straight-forward substitution of the respective time-dependent 
functions into the NS, CS, and SS results for the three 
$(TQ,~TM,~TO)$ systems.  In Section 5 we review this procedure.  [We 
refer the  interested reader to Refs  \cite{TMI,TMII} for 
the  equations necessary to implement the procedure, using the
functions  given in an  Appendix.]  

In Section 6, we give the 
expectation values for $\langle x \rangle$ and $\langle p \rangle$, 
and discuss the associated classical motions.  We also 
give the squeezed-state uncertainties and uncertainty products.  

For the condition $a \ge 0$ and $b\ge0$, 
Kim \cite{spk1} has partially studied the 
$TM$ system, 
obtaining the  number eigenstates in terms of 
implicit quantities.  The greater power of our formalism, which starts 
from a $TO$ equation, is that it allows a general calculation 
of the oscillator algebras, the number states, CS, SS, and 
expectation values for all three ($TQ,~TM,~TO$) Schr\"odinger 
systems.  This entails the complication of dealing 
with the various regimes for $a$ and $b$ in the transformation 
$TM \rightarrow TO$.  However, when dealing with $TM$, the solutions 
are clear.   
 

\section{HOW TO OBTAIN THE $TQ$ TIME-DEPENDENT FUNCTIONS}

\subsection{The Differential Equations for Cases 1 and 2}

As demonstrated in Refs. \cite{TMI,TMII} (also see Ref. \cite{paperII}), 
to obtain the time-dependent Lie 
symmetries for a $TO$ Schr\"odinger equation, we must solve the 
ordinary differential equation
\begin{equation}
\ddot{\gamma}+2g^{(2)}(t')\gamma=0,\label{e:ex2ls1}
\end{equation}
with the particular $g^{(2)}(t')$ being studied. First   
find two real solutions, $\gamma_1(t')$ and 
$\gamma_2(t')$, of 
\begin{equation}
W_{t'}(\gamma_1,\gamma_2) = 1. \label{W}
\end{equation}
(The subscript on the Wronskian indicates the variable of differentiation.)  
The Wronskian of the solutions of a second-order differential equation 
of the form (\ref{e:ex2ls1}) is always a constant {\cite{aec}} and 
we choose that constant to be unity.  The motivation for this will be 
made clear later.  We refer to this process as normalization with respect 
to the Wronskian (\ref{W}).

To complete the determination of the basis operators for the $TO$-$os(1)$ 
algebras, we then need to find complex solutions, $\xi(t')$ and  
$\bar{\xi}(t')$, to Eq. (\ref{e:ex2ls1})  
such that their Wronskian satisfies 
\begin{equation}
W_{t'}(\xi,\bar{\xi})=\xi\dot{\bar{\xi}}-\dot{\xi}\bar{\xi}=-i.
\label{e:ex2ls4wcs}
\end{equation}
This is accomplished by writing $\xi(t')$ in terms of the real solutions: 
\begin{equation}
\xi(t')=\sqrt{\lfrac{1}{2}}\left(\gamma_1(t')+i\gamma_2(t')\right).
\label{e:ex2ls4cs}
\end{equation}
The fact that  $\gamma_1$ and $\gamma_2$ satisfy 
Eq. (\ref{W}) guarantees that the complex  $\xi$ and 
$\bar{\xi}$ satisfy Eq. (\ref{e:ex2ls4wcs}).

Returning to the main cases, their differential 
equations (\ref{e:ex2ls1}) are:

{\bf Case 1:} $a=1$.  
Together, Eqs. (\ref{ex2ttfna=1}) and (\ref{e:ex2ls1}) yield
\begin{equation}
\ddot{\gamma}+\omega^2\exp\left[\frac{1+b}{t_o}(t'-t_o')\right]\gamma=0,
\label{e:ex2ls4.1}
\end{equation}
where $t'-t_o'\in [0,\infty)$.  Three classes of solutions arise depending 
on whether: $(b+1)~\{>,=,<\}~0$.  As an example, below we will give the 
derivation of the $TO$ functions for the class $b+1>0$.

{\bf Case 2:} $a \ne 1$. 
When we combine Eqs. (\ref{ex2pt12}) and (\ref{e:ex2ls1}), 
we obtain the differential equation 
\begin{equation}
\ddot{\gamma}+\omega^2\left[1+\frac{1-a}{t_o}(t'-t_o')
\right]^{\frac{a+b}{1-a}}\gamma=0,\label{e:ex2ls4.2}
\end{equation}
where $(t'-t_o')\in [0,\lfrac{t_o}{a-1})$ if $a\in (1,\infty)$ and 
$(t'-t_o')\in [0,\infty)$ if $a\in (-\infty,1)$.  

Note that when $a=0$, then $\nu=0$ and the unitary transformation 
(\ref{ex2nu})
from $TQ$ to $TM$ is the identity.  Further, the 
transformation from $TM$ to $TO$ is also unity, i.e.,  $t'=t$.  
Thus, here  we are already dealing with a $TM$-type 
equation and so will not consider $a=0$ further.  


\subsection{Classification of all the Solutions}
 
To review, for this problem the 
$TO$ Schr\"odinger equation for a particular system depends upon the 
powers $a$ and $b$ in the Hamiltonian.  The problem naturally divides into 
two cases: Case 1 for which $a=1$ and Case 2 for which $a\ne 1$.  Once 
we have identified which Case the $TO$ equation belongs to, further 
division into classes 
depends upon the power $b$.  In our system notation, we 
denote this by the first two entries in a symbol $\{a;b;\}$.  

For Case 1 ($a=1$), 
we identified three classes of solutions depending upon 
the value or range of values for the power $b+1$:
$(b+1)~\{>,=,<\}~0$.  
We have listed these classes of solutions in 
the first column of Table 1.  We sometimes use the union 
\begin{equation}
\{1;\ne -1;\}=\{1;(-\infty,-1);\}\cup \{1;(-1,\infty);\},
\label{union1}
\end{equation}
to indicate that a particular result holds for 
both subsystems $\{1;(-\infty,-1);\}$ and $\{1;(-1,\infty);\}$.

For Case 2 ($a\ne 1$), we identify three classes of systems in
solving Eq. (\ref{e:ex2ls4.2}).  Given  $a$, the


\begin{center}

\begin{tabular}{|c|c||c|}
\multicolumn{3}{l}{Table 1. System notation for this problem. The symbol 
$\pm$ refers to the }\\
\multicolumn{3}{l}{sign of $1-a$}\\
\hline\hline
\multicolumn{2}{|c||}{}     &              \\*[-3mm]
\multicolumn{2}{|c||}{$TO$} & $TM$ and $TQ$\\*[2mm]\hline 
 & & \\*[-3mm] 
 Case 1                   &  Case 2                &  \\
$\{a;b;\}$                &  $\{a;b;\}$            &  $\{a;b;\}$ 
\\*[2mm]\hline
  & & \\*[-3mm]
$\{1;(-1,\infty);\}$      &  $\{\ne 1;(a-2,\infty);\}$   
& $\{\ne 0;(a-2,\infty);\}$\\*[2mm]
$\{1;(-\infty,-1);\}$     &  $\{\ne 1;(-\infty,a-2);\}$   
& $\{\ne 0;(-\infty,a-2);\}$ \\*[2mm]
$\{1;-1;\}$    &  & $\{1;-1\}$\\*[2mm]
 & $\{\ne 1;a-2;t_o<\frac{|1-a|}{2\omega};\pm;\}$ 
& $\{\ne 0,1;a-2;t_o<\frac{|1-a|}{2\omega};\pm;\}$\\*[2mm]
                   &  
$\{\ne 1;a-2;t_o=\frac{|1-a|}{2\omega};\pm;\}$ 
& $\{\ne 0,1;a-2;t_o=\frac{|1-a|}{2\omega};\pm;\}$\\*[2mm]
                   &  
$\{\ne 1;a-2;t_o>\frac{|1-a|}{2\omega};\pm;\}$ 
& $\{\ne 0,1;a-2;t_o>\frac{|1-a|}{2\omega};\pm;\}$\\*[2mm]
\hline\hline
\end{tabular}

\end{center}


\noindent value 
of $b$ falls into one of three possible classes: $b~\{>,=,<\}~(a-2)$.  
The third class in Table 1, corresponding to $b=a-2$, can be further 
partitioned into 
six subclasses depending upon the relationship $\{>,=,<\}$ between $t_o$ and 
$\frac{|1-a|}{2\omega}$ and the sign of $1-a$.  
Therefore, in our notation, we place at the end  
other conditions or comments as needed to specify the 
system.  For example, we  write 
$\{\ne 1;a-2;t_o<\frac{|1-a|}{2\omega};\pm;\}$, where $\pm$ 
refers to the sign of $1-a$.  
In the second column of Table 1, we list the notation for the 8 possible 
systems of solutions for Case 2.  Occasionally, we use the union
\begin{equation}
\{\ne 1;\ne a-2;\}=\{\ne 1;(-\infty,a-2);\}\cup \{\ne 1;(a-2,\infty);\},
\label{union2}
\end{equation}
to indicate that a result applies to both subsystems 
$\{\ne 1;(-\infty,a-2);\}$ and $\{\ne 1;(a-2,\infty);\}$.
[Appendix A of Ref. \cite{TMII} is useful in 
obtaining some of the  solutions of Case 2.  Appendix A of this paper 
gives some important properties of Bessel functions.]


\section{THE TIME-DEPENDENT FUNCTIONS}

\subsection{The $TO$ Functions}

In Appendix B, Table B-1, we  list, according to the classification 
scheme of Table 1, the complex solutions, $\xi(t')$, 
and their derivatives, $\dot{\xi}(t')$. 
This is done for all the classes and subclasses we have 
described; i.e., for each $TO$ system described by 
a Schr\"odinger equation of the type (\ref{ex2S3}).  The recursion 
relations listed in Appendix A are useful in the construction of the 
derivatives, $\dot{\xi}$.  In addition, in Table B-2, we give the real 
functions $\phi_3(t')$ [see Eq. (\cite{TMI}-21)] and 
its derivatives $\dot{\phi}_3(t')$ 
and $\ddot{\phi}_3(t')$.  The solution $\bar{\xi}(t')$ and its 
derivative $\dot{\bar{\xi}}(t')$ can be obtained from $\xi(t')$ and 
$\dot{\xi}(t')$ by complex conjugation.  These are all the functions 
that are required to specify the operators that form a basis for the 
oscillator algebra, $os(1)$, associated with Eq. (\ref{ex2S3}).  


{\bf $TO$ Functions for a Specific Class: $\{a;b;\} = 
\{1;b+1 >  0;\}$:} 
Now we give a demonstration of the 
method for a specific class of Case 1: 
$\{a=1; b+1>0\}$.  [There are also separate classes $\{a=1; b+1=0\}$
and $\{a=1; b+1<0\}$.]
To begin,  set  
\begin{equation}
w(s)=\gamma(t')~~~~~~~~~~~
\sigma=\frac{2\omega t_o}{|b+1|}\exp\left[\frac{b+1}{2t_o}(t'-t_o')\right].
\label{e:ex2ls4.1a1} 
\end{equation}
Eq. (\ref{e:ex2ls4.1}) can then be transformed into  
\begin{equation}
\sigma^2\frac{d^2w}{d\sigma^2}+\sigma\frac{dw}{d\sigma}+\sigma^2w=0,
\label{e:ex2ls4.1a}
\end{equation}
which is Bessel's equation with zero eigenvalue.  It has real solutions
\begin{equation}
w_1(\sigma)=J_0(\sigma),~~~~~~w_2(\sigma)=Y_0(\sigma),\label{e:ex2ls4.1ars}
\end{equation}
where $J_n$ and $Y_n$ are Bessel functions of the first and second kind.
(See Appendix A.)  
When $b+1 > 0$, the variable $\sigma\in[\frac{2\omega t_o}{|b+1|},\infty)$.

We take, for the real solutions of Eq. (\ref{e:ex2ls4.1}), 
\begin{equation}
\gamma_1(t')=C_1J_0(\sigma),~~~~~~\gamma_2(t')=C_2Y_0(\sigma).
\label{e:ex2ls4.1rs}
\end{equation}
The constants $C_1$ and $C_2$ are chosen so that the Wronskian is 
\begin{equation}
W_{t'}(\gamma_1,\gamma_2)
=\gamma_1\dot{\gamma}_2-\dot{\gamma}_1\gamma_2 
=  C_1C_2\frac{d\sigma}{dt'}W_{\sigma}(J_0,Y_0)
  =  C_1C_2\left(\frac{b+1}{\pi t_o}\right)
  =  1.\label{e:ex2ls1w}
\end{equation}
[For the Wronskian $W_{\sigma}(J_0,Y_0)$, see Eq. (\ref{appa1}) 
in Appendix A.]
Taking $C_1=C_2$, we obtain 
\begin{equation}
C_1=\sqrt{\frac{\pi t_o}{b+1}}=C_2,\label{e:ex2ls4.1nc1}
\end{equation}
\begin{equation}
\gamma_1(t')=\sqrt{\lfrac{\pi t_o}{|b+1|}}J_0(\sigma),~~~~~~
\gamma_2(t')=\sqrt{\lfrac{\pi t_o}{|b+1|}}Y_0(\sigma),\label{e:ex2ls4.1rs1}
\end{equation}
where we have written $|b+1|$ for convenience.  

A complex solution is  obtained from Eqs. (\ref{e:ex2ls4.1rs1}) and 
(\ref{e:ex2ls4cs}):
\begin{equation}
\xi(t')=\sqrt{\lfrac{\pi t_o}{2|b+1|}}\left(J_0(\sigma)+iY_0(\sigma)\right)
=\sqrt{\lfrac{\pi t_o}{2|b+1|}}H_0^{(1)}(\sigma),\label{e:ex2ls4.1cs1}
\end{equation}
where $H_0^{(1)}(\sigma)$ is a Hankel function.  When the argument of a 
Hankel function is real, 
\begin{equation}
H_{j}^{(2)}=\bar{H}_{j}^{(1)},\label{e:ex2lsH}
\end{equation}
for all real $j$.  Hence, we write
\begin{equation}
\bar{\xi}(t')=\sqrt{\lfrac{\pi t_o}{2|b+1|}}\bar{H}_0^{(1)}(\sigma).
\label{e:ex2ls4.1cs1cc}
\end{equation}
[Eq. (\ref{appa4})  in Appendix A is helpful in establishing that Eqs. 
(\ref{e:ex2ls4.1cs1})  and (\ref{e:ex2ls4.1cs1cc})
satisfy Eq. (\ref{e:ex2ls4wcs}).] 
We can obtain $\dot{\xi}(t')$ from  $\xi(t')$. 
Then from Eq. (\cite{TMI}-21)
we have  the functions, $\phi_j$, $j=1,2,3$, 
\begin{equation}
\phi_1(t')=\xi^2(t'),~~~~\phi_2(t')=\bar{\xi}^2(t'),~~~~\phi_3(t') 
= 2\xi(t')\bar{\xi}(t').\label{phi}
\end{equation}
This gives us the $TO$ solutions for this particular class.  


{\bf A Special System of $TO$ Functions: 
$\{a;b;\} = \{\ne 1;-a;\}$:} 
The systems $\{\ne 1;-a;\}$ are of special interest.  When 
$a\in (-\infty,1)$, we have the  systems $\{(-\infty,1);-a;\} \subset
\{\ne 1;(a-2,\infty);\}$  and when $a\in (1,\infty)$, we have 
$\{(1,\infty);-a;\}\subset \{\ne 1;(-\infty,a-2);\}$.  
Therefore,  
the complex solution, $\xi(t')$, can be read from Table B-1:
\begin{eqnarray}
\{(-\infty,1);-a;\}:&~&\hspace{.7in}
\xi(t')=\sqrt{\frac{\pi t_o}{4|1-a|}}\sqrt{v}H_{\frac{1}{2}}^{(1)}(\tau),
\label{sc-a1}  \\
\{(1,\infty);-a;\}:~&~&\hspace{.7in}
\xi(t')=\sqrt{\frac{\pi t_o}{4|1-a|}}\sqrt{v}
\bar{H}_{\frac{1}{2}}^{(1)}(\tau),
\label{sc-a2}  \\
H_{\frac{1}{2}}^{(1)}(\tau)=-i\left(\frac{2}{\pi\tau}\right)^{\frac{1}{2}}
e^{i\tau},
&~&~~~~~~\bar{H}_{\frac{1}{2}}^{(1)}(\tau)=i\left(\frac{2}{\pi\tau}
\right)^{\frac{1}{2}}e^{-i\tau},~~~~~~~
\tau=\lfrac{\omega t_o}{|1-a|}v.\label{sc-a4}
\end{eqnarray}

We can perform our calculations with these functions for the systems 
$\{\ne 1;-a;\}$ or it is more constructive to observe that when $b=-a$, 
the differential equation (\ref{e:ex2ls4.2}) reduces to a 
simple harmonic oscillator equation 
which has the complex solutions 
\begin{equation}
\xi(t')=\sqrt{\lfrac{1}{2\omega}}e^{i\omega(t'-t_o')},\label{sc-a12}
\end{equation}
and $\bar{\xi}(t')$, the same solutions we use for the system $\{1;-1\}$.
The solutions (\ref{sc-a1}) and (\ref{sc-a2}) reduce to Eq. 
(\ref{sc-a12}) up to a complex constant of modulus 1.  It is important 
to recognize that normalization with respect to the Wronskian 
(\ref{e:ex2ls4wcs}) can only be carried out up to a complex constant 
of modulus 1.  We prefer to use the function (\ref{sc-a12}) and its 
complex conjugate. 
 
The real function $\phi_3(t')$, and its derivatives, $\dot{\phi}_3(t')$ 
and $\ddot{\phi}_3(t')$, are 
\begin{equation}
\phi_3(t')=\lfrac{1}{\omega},~~~~~~\dot{\phi}_3(t')=\ddot{\phi}_3(t')=0.
\label{sc-a16}
\end{equation}

See Table B-2 for the system $\{1;-1;\}$.


\subsection{The $TM$ Functions}

For $TM$ and $TQ$ systems, the distinction between Case 1 and Case 2 
is no longer appropriate. 
So,  there is  a new set of ranges for the powers $a$ and $b$ that 
define the $TM$ classes and subclasses
The $TM$ functions for the 6 classes of $TM$ systems and their subclasses 
can be obtained from the composition of the appropriate $TO$-functions 
with the mapping $t'(t)$ in Eqs. (\ref{ex2ttfna=1}) for $a=1$ and 
(\ref{ex2pt12}) for $a\ne 1$.  [See Eqs. (\cite{paperI}-95,96).]  

The parameterization of the $TM$ classes and subclasses is
shown in the last column of Table 1.  
[When it is not clear from the context, we shall always prefix 
the system designations in Table 1 with the symbol indicating which of 
the three Schr\"odinger equations we are dealing with.  For example,
$TO$-$\{1;-1;\}$ and $TM$-$\{1;-1;\}$.]

The end results are  
the functions $\hat{\xi}(t)$ and $\hat{\dot{\xi}}(t)$  
given in Appendix B, 
Table B-3, and  the functions $\hat{\phi}_3(t)$, $\hat{\dot{\phi}}_3(t)$, 
and $\hat{\ddot{\phi}}_3(t)$ given in Table B-4.


{\bf The Special System of (now) $TM$ Functions: 
$\{a;b;\} = \{\ne 1;-a;\}$:} 
For the systems $\{\ne 1;-a;\}$, we have 
\begin{eqnarray}
\hat{\xi}(t) = \sqrt{\lfrac{1}{2\omega}}e^{i\frac{\omega t_o}{1-a}
\left[\left(\frac{t}{t_o}\right)^{1-a}-1\right]},
&~~~~&
\hat{\dot{\xi}}(t)=i\omega\sqrt{\lfrac{1}{2\omega}}
e^{i\frac{\omega t_o}{1-a}\left[\left(\frac{t}{t_o}\right)^{1-a}-1\right]},
\label{sc-atm1} \\
\hat{\phi}_3(t)=\lfrac{1}{\omega},&~~~~~~&\hat{\dot{\phi}}_3(t)=
\hat{\ddot{\phi}}_3(t)=0.\label{sc-atm2}
\end{eqnarray}
Compare to the functions for the system $\{1;-1;\}$ in Tables B-3 and B-I4.


\subsection{The $TQ$ Functions}

Recall, as listed in Table 1, that the $TQ$ functions have the 
same class structure as the $TM$ functions.  These 
$TQ$ functions can be computed from the $TM$ functions 
using Eqs. (\cite{TMI}-41) to (\cite{TMI}-44).  
In Appendix B, Table B-5, we have listed  
$\Xi_P$ and $\Xi_X$.  In Table B-6, we give the coefficients $C_{3,T}$, 
$C_{3,D}$, and $C_{3,X^2}$ for each of the 6 classes of systems under 
consideration.  


{\bf The Special System of (now) $TQ$ Functions: 
$\{a;b;\} = \{\ne 1;-a;\}$:} 
For the systems $\{\ne 1;-a;\}$, the functions $\Xi_P$ and $\Xi_X$ are  
\begin{equation}
\Xi_P(t)=\sqrt{\lfrac{1}{2\omega}}\left(\frac{t}{t_o}\right)^{\frac{a}{2}}
e^{i\frac{\omega t_o}{1-a}\left[\left(\frac{t}{t_o}\right)^{1-a}-1\right]},
~~~~~~\Xi_X(t)=i\omega\sqrt{\lfrac{1}{2\omega}}\left(\frac{t}{t_o}\right)^{
-\frac{a}{2}}e^{i\frac{\omega t_o}{1-a}\left[\left(\frac{t}{t_o}
\right)^{1-a}-1\right]},\label{sc-atq1}
\end{equation}
while the coefficients $C_{3,T}$, $C_{3,D}$, and $C_{3,X^2}$ are
\begin{equation}
C_{3,T}=\lfrac{1}{\omega}\left(\frac{t}{t_o}
\right)^a,~~~~~~C_{3,D}=\frac{a}{2\omega t_o}\left(\frac{t}{t_o}
\right)^{a-1},~~~~~~C_{3,X^2}=0.\label{sc-atq4}
\end{equation}

\subsection{Initial Values of the Functions}

In subsequent calculations, we shall need initial values for time-dependent 
functions.  When $t'=t'_o$ or $t=t_o$, we identify the initial value of the 
function by a superscript $^o$, such as $\xi^o$, for example.  
Recall that for this paper, we have assumed that $t_o>0$.


\section{THE OSCILLATOR SUBALGEBRAS}

For each class of system, described by 
the $TM$ Schr\"odinger equation (\ref{ex2S2}) 
and its related $TQ$ and $TO$ Schr\"odinger equations, (\ref{ex2S1}) 
and (\ref{ex2S3}), respectively, there is an oscillator algebra.  All 
these oscillator algebras are isomorphic.  

For $TO$  systems, 
one obtains the operators by substituting 
the functions of Tables B-1 and B-2 into 
Eq. (\cite{TMI}-18)  for  $J_{3\pm}$, and into (\cite{TMI}-17)  
for $M_3$.  For $TM$ systems, 
one obtains the operators by substituting   
the functions of Tables B-3 and B-4 into Eq. (\cite{TMI}-31)  for 
$\hat{J}_{2\pm}$, and into  Eq. (\cite{TMI}-30)  for $\hat{M}_2$. 
For $TQ$ systems, 
one obtains the operators by substituting the   
the functions of Tables B-5 and B-6 into  
Eq. (\cite{TMI}-40)  for $J_{1\pm}$ and  $M_1$.


{\bf The $os(1)$ Operators for $TM$-$\{\ne 0;(a-2,\infty);\}$:} 
Because of the large number of possible systems, we cite only one example, 
the $os(1)$ basis operators for for $TM$-$\{\ne 0;(a-2,\infty);\}$.  These 
operators are obtained by substituting the appropriate functions 
in Tables B-3 and B-4 (with $\tau$ defined there) 
into Eqs. (\cite{paperII}-33,38,46-52),
\begin{eqnarray}
\hat{M}_2 & = & \lfrac{\pi t_o}{|b-a+2|}\left(\frac{t}{t_o}\right)
H_{\frac{1}{q}}^{(1)}(\tau)\bar{H}_{\frac{1}{q}}^{(1)}(\tau)T\nonumber\\
 & & -\lfrac{\pi}{4}\tau\left[H_{\frac{1}{q}-1}^{(1)}(\tau)
\bar{H}_{\frac{1}{q}}^{(1)}(\tau)+H_{\frac{1}{q}}^{(1)}(\tau)
\bar{H}_{\frac{1}{q}-1}^{(1)}(\tau)\right]D\nonumber\\
 & & +\lfrac{\pi|b-a+2|}{8t_o}\left(\frac{t}{t_o}\right)^{a-1}\tau^2
\left[H_{\frac{1}{q}-1}^{(1)}(\tau)\bar{H}_{\frac{1}{q}-1}^{(1)}(\tau)
-H_{\frac{1}{q}}^{(1)}(\tau)\bar{H}_{\frac{1}{q}}^{(1)}(\tau)\right]X^2,
\label{genex1} \\
\hat{J}_{2-} & = & i\sqrt{\lfrac{\pi t_o}{2|b-a+2|}}\left(\frac{t}{t_o}
\right)^{\frac{1-a}{2}}\left\{H_{\frac{1}{q}}^{(1)}(\tau)P
-\lfrac{|b-a+2|}{2t_o}\left(\frac{t}{t_o}\right)^{a-1}
H_{\frac{1}{q}-1}^{(1)}(\tau)X\right\},
\label{j2plus} \\*[2mm]
\hat{J}_{2+} & = & i\sqrt{\lfrac{\pi t_o}{2|b-a+2|}}\left(\frac{t}{t_o}
\right)^{\frac{1-a}{2}}\left\{-\bar{H}_{\frac{1}{q}}^{(1)}(\tau)P
+\lfrac{|b-a+2|}{2t_o}\left(\frac{t}{t_o}\right)^{a-1}
\bar{H}_{\frac{1}{q}-1}^{(1)}(\tau)X\right\}.  \label{genex2}
\end{eqnarray}
One can show that these generators satisfy the $os(1)$ Lie algebra
commutation relations 
\begin{equation}
[\hat{M}_2,\hat{J}_{2\pm}]=\pm \hat{J}_{2\pm},~~~~~~
[\hat{J}_{2-},\hat{J}_{2+}]=I.  \label{os1tmex}
\end{equation}


\section{Wave Functions}

For time-dependent systems of this type, 
explicit space-time representations of the wave functions are not necessary 
for the computation of expectation values for position and momentum, 
their uncertainties and uncertainty products \cite{paperII}.  Nevertheless, 
they are of independent interest and number-state, coherent-state, and 
squeezed-state wave functions may be constructed for our problem  from our 
results in Ref. \cite{TMI} and the tables in Appendix B.
 
{\bf Number States:} 
In Ref. \cite{TMI} we refer the reader to Eqs. (\cite{TMI}-51) 
for $TO$ number states, (\cite{TMI}-58) for $TM$ number states, 
and (\cite{TMI}-63) for $TQ$ number states.  
 
{\bf Coherent States:} 
Eqs. (\cite{TMI}-66), (\cite{TMI}-68), and (\cite{TMI}-74) are the 
respective coherent states for $TO$, $TM$, and $TQ$ systems.

{\bf Squeezed States:} 
For squeezed states, we have Eqs. (\cite{TMI}-91) for $TO$ systems, 
(\cite{TMI}-96) for $TM$ systems, and (\cite{TMI}-103) for $TQ$ systems.


\section{EXPECTATION VALUES}

\subsection{The Dynamical Variables $\langle x \rangle$ and 
$\langle p \rangle$}

Elsewhere,  formulas were obtained for 
the expectation values of position and momentum from 
time-dependent quadratic Hamiltonians for general 
$TQ$, $TM$, and $TO$ systems, respectively, in Eqs. 
(86), (87) and (88) of Ref. \cite{paperII}, respectively.  
These results were given in 
terms of general time-dependent functions of the type discussed 
in the present Section 2.  

In particular, for our present problem 
the time-dependent $\xi$ and other functions are given in Appendix B.
[Note:  For the $TQ$ system, the results are given in terms of the 
$\Xi_P(t)=\hat{\xi}(t)e^{\nu}$ and $\Xi_X(t)=\hat{\xi}(t)e^{-\nu}$ 
of Eq. (\cite{TMI}-41).]
Therefore, depending on the complicated various regimes of the 
parameters $a$, $b$, $\omega$, and $t_o$, we can use 
Tables B-1, B-3, and B-5, to obtain $\langle x \rangle$ and 
$\langle p \rangle$ for the $TO$, $TM$, and $TQ$ systems, 
respectively.  The results of our computations of 
$\langle x \rangle$ and $\langle p \rangle$
for all the various (sub)systems, are displayed in Tables 2 to 7.  


\subsection{The Classical Motion}

As before, the coherent-state and squeezed-state 
expectation values $\langle x \rangle$ and 
$\langle p \rangle$ should obey the classical Hamiltonian 
equations of motion:
\begin{equation}
\dot{x} = \frac{\partial H}{\partial p}, ~~~~~~~~~~
\dot{p} =-\frac{\partial H}{\partial x}.   \label{ceqm}
\end{equation}

Now, the classical Hamiltonians 
associated with our Schr\"odinger equations are
\begin{eqnarray}
TO:~~~~~H&=& \lfrac{1}{2}p^2 + g^{(2)}(t'){x^2},
 \\
TM:~~~~~\hat{H}&=& \frac{1}{2}\left(\frac{t_o}{t}\right)^a p^2
          +\frac{1}{2}{\omega^2}\left(\frac{t}{t_o}\right)^b{x^2},
\\
TQ:~~~~~H&=&\frac{1}{2}p^2 +\frac{a}{2t}xp
     +\frac{1}{2}{\omega^2}\left(\frac{t}{t_o}\right)^{b-a}  {x^2},
\end{eqnarray}
\noindent where $g^{(2)}(t')$ is given in 
Eqs. (\ref{ex2ttfna=1}) and (\ref{ex2pt12}).
Applying these to the classical equations of motion (\ref{ceqm})  
one obtains (``dot" is $d/dt'$ or $d/dt$ as appropriate)


\begin{center}

\begin{tabular}{|r|c|}
\multicolumn{2}{l}{Table 2. $\langle x(t')\rangle$ for the $TO$ systems. 
Variables and parameters are defined }\\
\multicolumn{2}{l}{as follows: 
$\sigma=\frac{2\omega t_o}{|b+1|}\exp\left[\frac{b+1}{2t_o}(t'-t_o')\right]$;  
$v=1+\frac{1-a}{t_o}(t'-t_o')$; $\tau=\frac{2\omega t_o}{|b-a+2|}v^{q/2}$; }\\
\multicolumn{2}{l}{$\sigma_o=\frac{2\omega t_o}{|b+1|}$; 
$\tau_o=\frac{2\omega t_o}{|b-a+2|}$; $q=\frac{b-a+2}{1-a}$, and 
$\Delta^2=\left|1-\frac{4\omega^2t_o^2}{(1-a)^2}\right|$.}\\*[1.5mm]
\hline\hline
 & \\*[-4mm]
\multicolumn{1}{|c|}{System} & $\langle x(t')\rangle$ \\*[1mm]\hline
      & \\*[-3mm]
$\scsz{\{1;\ne -1;\}}$ & $\scsz{p_o\frac{\pi t_o}{b+1}\left[
Y_0(\sigma)J_0(\sigma_o)-J_0(\sigma)Y_0(\sigma_o)\right]+x_o
\frac{\pi\sigma_o}{2}\left[J_0(\sigma)Y_{-1}(\sigma_o)
-Y_0(\sigma)J_{-1}(\sigma_o)\right]}$\\*[6mm]\hline
 & \\*[-3mm] 
$\scsz{\{1;-1;\}}$ & $\scsz{p_o\frac{1}{\omega}
\sin{\omega(t'-t_o')}+x_o\cos{\omega(t'-t_o')}}$ \\*[3mm]\hline
 & \\*[-3mm]
$\scsz{\{\ne 1;\ne a-2;\}}$
& $\begin{array}{l}\scsz{p_o\frac{\pi t_o}{b-a+2}\sqrt{v}\left[Y_{\frac{1}{q}}(\tau)
J_{\frac{1}{q}}(\tau_o)-J_{\frac{1}{q}}(\tau)Y_{\frac{1}{q}}(\tau_o)\right]}
\\
\hspace{1cm}\scsz{+x_o\sqrt{v}\,\frac{\pi\tau_o}{2}\left[J_{\frac{1}{q}}(\tau)Y_{\frac{1}{q}-1}
(\tau_o)-Y_{\frac{1}{q}}(\tau)J_{\frac{1}{q}-1}(\tau_o)\right]}\end{array}$
\\*[6mm]
\hline
 & \\*[-3mm]
$\scsz{\{\ne 1;a-2;t_o<\frac{|1-a|}{2\omega};\}}$
& $\scsz{p_o\frac{2t_o}
{(1-a)\Delta}\sqrt{v}\sinh{\left(\frac{\Delta}{2}\ln{v}\right)}+
x_o\frac{1}{\Delta}\sqrt{v}\left[\Delta\cosh{\left(\frac{\Delta}{2}\ln{v}
\right)}-\sinh{\left(\frac{\Delta}{2}\ln{v}\right)}\right]}$\\*[5mm]
$\scsz{\{\ne 1;a-2;t_o=\frac{|1-a|}{2\omega};\}}$
& $\scsz{p_o\frac{t_o}{1-a}
\sqrt{v}\ln{v}+x_o\sqrt{v}\left(1-\frac{1}{2}\ln{v}\right)}$ \\*[5mm]
$\scsz{\{\ne 1;a-2;t_o>\frac{|1-a|}{2\omega};\}}$
& $\scsz{p_o\frac{2t_o}
{(1-a)\Delta}\sqrt{v}\sin{\left(\frac{\Delta}{2}\ln{v}\right)}+
x_o\frac{1}{\Delta}\sqrt{v}\left[\Delta\cos{\left(\frac{\Delta}{2}\ln{v}\right)}
-\sin{\left(\frac{\Delta}{2}\ln{v}\right)}
\right]}$ \\*[5mm]\hline\hline
\end{tabular}

\end{center}

\begin{center}

\begin{tabular}{|r|c|}
\multicolumn{2}{l}{Table 3. $\langle p(t')\rangle$ for the $TO$ systems.
Variables and parameters are defined }\\
\multicolumn{2}{l}{as follows:
$\sigma=\frac{2\omega t_o}{|b+1|}\exp\left[\frac{b+1}{2t_o}(t'-t_o')
\right]$; $v=1+\frac{1-a}{t_o}(t'-t_o')$; $\tau=\frac{2\omega t_o}{|b-a+2|}
v^{q/2}$; }\\ 
\multicolumn{2}{l}{$\sigma_o=\frac{2\omega t_o}{|b+1|}$;
$\tau_o=\frac{2\omega t_o}{|b-a+2|}$; $q=\frac{b-a+2}{1-a}$, and 
$\Delta^2=\left|1-\frac{4\omega^2t_o^2}{(1-a)^2}\right|$.}\\*[1.5mm]
\hline\hline
  & \\*[-4mm]
\multicolumn{1}{|c|}{System} & $\langle p(t')\rangle$ \\*[1mm]\hline
  & \\*[-3mm]
$\scsz{\{1;\ne -1;\}}$ & $\begin{array}{l}\scsz{p_o\frac{\pi\sigma}{2}
\left[Y_{-1}(\sigma)J_0(\sigma_o)-J_{-1}(\sigma)Y_0(\sigma_o)\right]}\\
\scsz{\hspace{1cm}+x_o\frac{\pi(b+1)}{4t_o}\sigma\sigma_o\left[
J_{-1}(\sigma)Y_{-1}(\sigma_o)-Y_{-1}(\sigma)J_{-1}(\sigma_o)\right]} 
\end{array}$\\*[6mm]\hline
  & \\*[-3mm]
$\scsz{\{1;-1;\}}$ & $\scsz{p_o\cos{\omega(t'-t_o')}
-x_o\omega\sin{\omega(t'-t_o')}}$\\*[3mm]\hline
 & \\*[-3mm]
$\scsz{\{\ne1;\ne a-2;\}}$ & $\begin{array}{l} \scsz{p_o\frac{1}{\sqrt{v}}
\frac{\pi\tau}{2} \left[Y_{\frac{1}{q}-1}(\tau)J_{\frac{1}{q}}(\tau_o)-
J_{\frac{1}{q}-1}(\tau)Y_{\frac{1}{q}}(\tau_o)\right]}\\
\scsz{\hspace{1cm}+x_o\frac{\pi(b-a+2)}{4t_o}\frac{1}{\sqrt{v}}\tau\tau_o
\left[J_{\frac{1}{q}-1}(\tau)Y_{\frac{1}{q}-1}(\tau_o)-Y_{\frac{1}{q}-1}
(\tau)J_{\frac{1}{q}-1}(\tau_o)\right]} \end{array}$\\*[9mm]\hline
 & \\*[-3mm]
$\scsz{\{\ne 1;a-2;t_o<\frac{|1-a|}{2\omega};\}}$ & $\scsz{p_o
\frac{1}{\Delta}\frac{1}{\sqrt{v}}\left[\Delta\cosh{\left(
\frac{\Delta}{2}\ln{v}\right)}+\sinh{\left(\frac{\Delta}{2}\ln{v}\right)}
\right]-x_o\frac{2\omega^2t_o}
{(1-a)\Delta}\frac{1}{\sqrt{v}}\sinh{\left(\frac{\Delta}{2}\ln{v}\right)}}$ 
\\*[5mm]
$\scsz{\{\ne 1;a-2;t_o=\frac{|1-a|}{2\omega};\}}$ & $\scsz{p_o
\frac{1}{\sqrt{v}}
\left(1+\frac{1}{2}\ln{v}\right)-x_o\frac{\omega^2t_o}{(1-a)}
\frac{1}{\sqrt{v}}\ln{v}}$\\*[5mm]
$\scsz{\{\ne 1;a-2;t_o>\frac{|1-a|}{2\omega};\}}$ & $\scsz{p_o
\frac{1}{\Delta}\frac{1}{\sqrt{v}}\left[\Delta\cos{\left(
\frac{\Delta}{2}\ln{v}\right)}+\sin{\left(\frac{\Delta}{2}\ln{v}\right)}
\right]-x_o\frac{2\omega^2t_o}{(1-a)\Delta}\frac{1}{\sqrt{v}}
\sin{\left(\frac{\Delta}{2}\ln{v}\right)}}$
\\*[5mm]\hline\hline
\end{tabular}

\end{center}

\newpage

\begin{center}

\begin{tabular}{|r|c|}
\multicolumn{2}{l} {Table 4. $\langle x(t)\rangle$ for the $TM$ systems. 
Variables and parameters are defined }\\
\multicolumn{2}{l}{as follows: 
$\tau=\frac{2\omega t_o}{|b-a+2|}\left(\frac{t}{t_o}\right)^{
\frac{b-a+2}{2}}$; $\chi=\frac{1}{2}(1-a)\ln{\frac{t}{t_o}}$; 
$\tau_o=\frac{2\omega t_o}{|b-a+2|}$; $q=\frac{b-a+2}{1-a}$,}\\
\multicolumn{2}{l}{and $\Delta^2=\left|1-\frac{4\omega^2t_o^2}{(1-a)^2}
\right|$.}\\*[1.5mm]
\hline\hline
  & \\*[-4mm]
\multicolumn{1}{|c|}{System} & $\langle x(t')\rangle$ \\*[1mm]\hline
  & \\*[-3mm]
$\scsz{\{\ne 0;\ne a-2;\}}$ & $\begin{array}{l}\scsz{p_o\frac{\pi t_o}
{b-a+2}\left(\frac{t}{t_o}\right)^{\frac{1-a}{2}}\left[Y_{\frac{1}{q}}(\tau)
J_{\frac{1}{q}}(\tau_o)-J_{\frac{1}{q}}(\tau)Y_{\frac{1}{q}}(\tau_o)\right]}\\
\scsz{\hspace{1cm}+ x_o\left(\frac{t}{t_o}\right)^{\frac{1-a}{2}}
\frac{\pi\tau_o}{2}\left[J_{\frac{1}{q}}(\tau)Y_{\frac{1}{q}-1}(\tau_o)
-Y_{\frac{1}{q}}(\tau)J_{\frac{1}{q}-1}(\tau_o)\right]}\end{array}$
\\*[9mm]\hline
 & \\*[-3mm]
$\scsz{\{1,-1\}}$ & $\scsz{p_o\frac{1}{\omega}\sin{\left(\omega t_o
\ln{\frac{t}{t_o}}\right)}+x_o\cos{\left(\omega t_o\ln{\frac{t}{t_o}}
\right)}}$\\*[3mm]\hline
 & \\*[-3mm]
$\scsz{\{\ne 0,1;a-2;t_o<\frac{|1-a|}{2\omega};\}}$ & $\scsz{p_o\frac{2t_o}
{(1-a)\Delta}\left(\frac{t}{t_o}\right)^{\frac{1-a}{2}}
\sinh{\left(\Delta\chi\right)}+x_o
\frac{1}{\Delta}\left(\frac{t}{t_o}\right)^{\frac{1-a}{2}}\left[
\Delta\cosh{\left(\Delta\chi\right)}-\sinh{\left(\Delta\chi\right)}\right]}$
\\*[5mm]
$\scsz{\{\ne 0,1;a-2;t_o=\frac{|1-a|}{2\omega};\}}$ & $\scsz{
p_o\frac{2t_o}{1-a}\left(\frac{t}{t_o}\right)^{\frac{1-a}{2}}\chi+
x_o\left(\frac{t}{t_o}\right)^{\frac{1-a}{2}}(1-\chi)}$ \\*[5mm]
$\scsz{\{\ne 0,1;a-2;t_o>\frac{|1-a|}{2\omega};\}}$ & $\scsz{
p_o\frac{2t_o}{(1-a)\Delta}\left(\frac{t}{t_o}\right)^{\frac{1-a}{2}}
\sin{\left(\Delta\chi\right)}+x_o\frac{1}{\Delta}\left(\frac{t}{t_o}
\right)^{\frac{1-a}{2}}\left[\Delta\cos{\left(\Delta\chi\right)}
-\sin{\left(\Delta\chi\right)}\right]}$ \\*[5mm]\hline\hline
\end{tabular}

\end{center}

\begin{center}

\begin{tabular}{|r|c|}
\multicolumn{2}{l} {Table V. $\langle p(t)\rangle$ for the $TM$ systems.
Variables and parameters are defined }\\
\multicolumn{2}{l}{as follows:  
$\tau=\frac{2\omega t_o}{|b-a+2|}\left(\frac{t}{t_o}\right)^{
\frac{b-a+2}{2}}$; $\chi=\frac{1}{2}(1-a)\ln{\frac{t}{t_o}}$; 
$\tau_o=\frac{2\omega t_o}{|b-a+2|}$; $q=\frac{b-a+2}{1-a}$,}\\ 
\multicolumn{2}{l}{and $\Delta^2=\left|1-\frac{4\omega^2t_o^2}{(1-a)^2}
\right|$.}\\*[1.5mm]
\hline\hline
  & \\*[-4mm]
\multicolumn{1}{|c|}{System} & $\langle p(t')\rangle$ \\*[1mm]\hline
  & \\*[-3mm]
$\scsz{\{\ne 0;\ne a-2\}}$ & $\begin{array}{l} \scsz{p_o\left(\frac{t}{t_o}
\right)^{\frac{a-1}{2}}\frac{\pi\tau}{2}\left[Y_{\frac{1}{q}-1}(\tau)
J_{\frac{1}{q}}(\tau_o)-J_{\frac{1}{q}-1}(\tau)Y_{\frac{1}{q}}(\tau_o)
\right]}\\
\scsz{\hspace{1cm}+x_o\frac{\pi(b-a+2)}{4t_o}\left(\frac{t}{t_o}
\right)^{\frac{a-1}{2}}\tau\tau_o\left[J_{\frac{1}{q}-1}(\tau)
Y_{\frac{1}{q}-1}(\tau_o)-Y_{\frac{1}{q}-1}(\tau)J_{\frac{1}{q}-1}(\tau_o)
\right]} \end{array}$
\\*[9mm]\hline
 & \\*[-3mm]
$\scsz{\{1;-1\}}$ & $\scsz{p_o\cos{\left(\omega t_o\ln{\frac{t}
{t_o}}\right)}-x_o\omega\sin{\left(\omega t_o\ln{\frac{t}{t_o}}\right)}}$
\\*[4mm]\hline
 & \\*[-3mm]
$\scsz{\{\ne 0,1;a-2;t_o<\frac{|1-a|}{2\omega}\}}$ & $\scsz{
p_o\frac{1}{\Delta}
\left(\frac{t}{t_o}\right)^{\frac{a-1}{2}}\left[\Delta\cosh{\left(\Delta\chi
\right)}+\sinh{\left(\Delta\chi\right)}\right]-x_o\frac{2\omega^2t_o}
{(1-a)\Delta}\left(\frac{t}{t_o}\right)^{\frac{a-1}{2}}\sinh{\left(\Delta\chi
\right)}}$\\*[5mm]
$\scsz{\{\ne 0,1;a-2;t_o<\frac{|1-a|}{2\omega}\}}$ & $\scsz{
p_o\left(\frac{t}{t_o}
\right)^{\frac{a-1}{2}}(1+\chi)-x_o\frac{\omega^2t_o}{1-a}\left(\frac{t}{t_o}
\right)^{\frac{a-1}{2}}\chi}$\\*[5mm]
$\scsz{\{\ne 0,1;a-2;t_o>\frac{|1-a|}{2\omega}\}}$ & $\scsz{
p_o\frac{1}{\Delta}
\left(\frac{t}{t_o}\right)^{\frac{a-1}{2}}\left[\Delta\cos{\left(\Delta\chi
\right)}+\sin{\left(\Delta\chi\right)}\right]-x_o\frac{2\omega^2t_o}
{(1-a)\Delta}\left(\frac{t}{t_o}\right)^{\frac{a-1}{2}}\sin{\left(\Delta\chi
\right)}}$\\*[5mm]\hline\hline
\end{tabular} 

\end{center}

\begin{center}

\begin{tabular}{|r|c|}
\multicolumn{2}{l} {Table 6. $\langle x(t)\rangle$ for the $TQ$ systems. 
Variables and parameters are defined }\\
\multicolumn{2}{l}{as follows:  
$\tau=\frac{2\omega t_o}{|b-a+2|}\left(\frac{t}{t_o}\right)^{
\frac{b-a+2}{2}}$; $\chi=\frac{1}{2}(1-a)\ln{\frac{t}{t_o}}$; 
$\tau_o=\frac{2\omega t_o}{|b-a+2|}$; $q=\frac{b-a+2}{1-a}$,}\\
\multicolumn{2}{l}{and $\Delta^2=\left|1-\frac{4\omega^2t_o^2}{(1-a)^2}
\right|$.}\\*[1.5mm]
\hline\hline
  & \\*[-4mm]
\multicolumn{1}{|c|}{System} & $\langle x(t')\rangle$ \\*[1mm]\hline
  & \\*[-3mm]
$\scsz{\{\ne 0;\ne a-2;\}}$ & $\begin{array}{l}\scsz{p_o
\frac{\pi t_o}{|b-a+2|}\left(\frac{t}{t_o}\right)^{\frac{1}{2}}\left[
Y_{\frac{1}{q}}(\tau)J_{\frac{1}{q}}(\tau_o)-J_{\frac{1}{q}}(\tau)
Y_{\frac{1}{q}}(\tau_o)\right]}\\
\scsz{\hspace{1cm}+x_o \left(\frac{t}{t_o}
\right)^{\frac{1}{2}}\frac{\pi\tau_o}{2}\left[J_{\frac{1}{q}}(\tau)
Y_{\frac{1}{q}-1}(\tau_o)-Y_{\frac{1}{q}}(\tau)J_{\frac{1}{q}-1}(\tau_o)
\right]}\end{array}$
\\*[9mm]\hline
 & \\*[-3mm]
$\scsz{\{1;-1;\}}$ & 
$\scsz{\frac{p_o}{\omega}\left(\frac{t}{t_o}\right)^{\frac{1}{2}}
\sin{\left(\omega t_o\ln{\frac{t}{t_o}}\right)}+x_o\left(\frac{t}{t_o}
\right)^{\frac{1}{2}}\cos{\left(\omega t_o\ln{\frac{t}{t_o}}\right)}}$
\\*[3mm]\hline
 & \\*[-3mm]
$\scsz{\{\ne 0,1;a-2;t_o<\frac{|1-a|}{2\omega};\}}$ & $\scsz{
p_o\frac{2t_o}{(1-a)\Delta}\left(\frac{t}{t_o}\right)^{\frac{1}{2}}
\sinh{\left(\Delta\chi\right)}+x_o
\frac{1}{\Delta}\left(\frac{t}{t_o}\right)^{\frac{1}{2}}\left[
\Delta\cosh{\left(\Delta\chi\right)}-\sinh{\left(\Delta\chi\right)}\right]}$
\\*[5mm]
$\scsz{\{\ne 0,1;a-2;t_o=\frac{|1-a|}{2\omega};\}}$ & $\scsz{
p_o\frac{2t_o}{1-a}\left(\frac{t}{t_o}\right)^{\frac{1}{2}}\chi+
x_o\left(\frac{t}{t_o}\right)^{\frac{1}{2}}(1-\chi)}$ 
\\*[5mm]
$\scsz{\{\ne 0,1;a-2;t_o>\frac{|1-a|}{2\omega};\}}$ & $\scsz{
p_o\frac{2t_o}{(1-a)\Delta}\left(\frac{t}{t_o}\right)^{\frac{1}{2}}
\sin{\left(\Delta\chi\right)}+x_o\frac{1}{\Delta}\left(\frac{t}{t_o}
\right)^{\frac{1}{2}}\left[\Delta\cos{\left(\Delta\chi\right)}
-\sin{\left(\Delta\chi\right)}\right]}$ \\*[5mm]\hline\hline
\end{tabular}

\end{center}

\begin{center}

\begin{tabular}{|r|c|}
\multicolumn{2}{l} {Table 7. $\langle p(t)\rangle$ for the $TQ$ systems.
Variables and parameters are defined }\\
\multicolumn{2}{l}{as follows:  
$\tau=\frac{2\omega t_o}{|b-a+2|}\left(\frac{t}{t_o}\right)^{
\frac{b-a+2}{2}}$; $\chi=\frac{1}{2}(1-a)\ln{\frac{t}{t_o}}$; 
$\tau_o=\frac{2\omega t_o}{|b-a+2|}$; $q=\frac{b-a+2}{1-a}$, }\\ 
\multicolumn{2}{l}{and $\Delta^2=\left|1-\frac{4\omega^2t_o^2}{(1-a)^2}
\right|$.}\\*[1.5mm]
\hline\hline
  & \\*[-4mm]
\multicolumn{1}{|c|}{System} & $\langle p(t')\rangle$ \\*[1mm]\hline
      & \\*[-3mm]
$\scsz{\{\ne 0;\ne a-2\}}$ &  $\begin{array}{l} \scsz{p_o
\left(\frac{t}{t_o}\right)^{-\frac{1}{2}}\frac{\pi\tau}{2}\left[
Y_{\frac{1}{q}-1}(\tau)J_{\frac{1}{q}}(\tau_o)-J_{\frac{1}{q}-1}(\tau)Y_{\frac{1}{q}}(\tau_o)
\right]}\\
\scsz{\hspace{1cm}+x_o\frac{\pi(b-a+2)}{4t_o}\left(\frac{t}{t_o}
\right)^{-\frac{1}{2}}\tau\tau_o\left[J_{\frac{1}{q}-1}(\tau)Y_{
\frac{1}{q}-1}(\tau_o)-Y_{\frac{1}{q}-1}(\tau)J_{\frac{1}{q}-1}
(\tau_o)\right]} \end{array}$
\\*[9mm]\hline
 & \\*[-3mm]
$\scsz{\{1;-1;\}}$ & $\scsz{p_o\left(\frac{t}{t_o}\right)^{-\frac{1}{2}}
\cos{\left(\omega t_o\ln{\frac{t}{t_o}}\right)}-x_o\omega\left(\frac{t}{t_o}
\right)^{-\frac{1}{2}}\sin{\left(\omega t_o\ln{\frac{t}{t_o}}\right)}}$
\\*[4mm]\hline
 & \\*[-3mm]
$\scsz{\{\ne 0,1;a-2;t_o<\frac{|1-a|}{2\omega};\}}$ & $\scsz{
p_o\frac{1}{\Delta}\left(\frac{t}{t_o}\right)^{-\frac{1}{2}}\left[
\Delta\cosh{\left(\Delta\chi
\right)}+\sinh{\left(\Delta\chi\right)}\right]-x_o\frac{2\omega^2t_o}
{(1-a)\Delta}\left(\frac{t}{t_o}\right)^{-\frac{1}{2}}\sinh{\left(\Delta\chi
\right)}}$\\*[5mm]
$\scsz{\{\ne 0,1;a-2;t_o=\frac{|1-a|}{2\omega};\}}$ & $\scsz{
p_o\left(\frac{t}{t_o}\right)^{-\frac{1}{2}}(1+\chi)-x_o
\frac{\omega^2 t_o}{1-a}\left(\frac{t}{t_o}
\right)^{-\frac{1}{2}}\chi}$\\*[5mm]
$\scsz{\{\ne 0,1;a-2;t_o>\frac{|1-a|}{2\omega};\}}$ & $\scsz{
p_o\frac{1}{\Delta}\left(\frac{t}{t_o}\right)^{-\frac{1}{2}}\left[
\Delta\cos{\left(\Delta\chi\right)}+\sin{\left(\Delta\chi\right)}\right]
-x_o\frac{2\omega^2t_o}{(1-a)\Delta}\left(\frac{t}{t_o}\right)^{
-\frac{1}{2}}\sin{\left(\Delta\chi\right)}}$\\*[5mm]\hline\hline
\end{tabular}
 
\end{center}

\begin{eqnarray}
&TO:&~~~~~~~\dot{ x }= p , ~~~~~~~~~~~~~~~~~~
\dot{ p } = -2g^{(2)}(t')x ,  \label{toem}
 \\
&TM:&~~~~~~~\dot{x }=\left(\frac{t_o}{t}\right)^a p , ~~~~~~~~~~~~~~
\dot{ p } = -{\omega^2}\left(\frac{t}{t_o}\right)^b x ,    \label{tmem}
\\
&TQ:&~~~~~~~\dot{x}= p 
                 + \lfrac{a}{2t} x ,~~~~~~~~
\dot{p} = -\omega^2 \left(\frac{t}{t_o}\right)^{b-a}  x 
               -\frac{a}{2t}p   \label{tqem}
\end{eqnarray}

The reader can verify that all the expectation values in Tables 2 to 
7 satisfy the equations of motion (\ref{toem}) - (\ref{tqem}). 
This specifically demonstrates 
the general results for quadratic time-dependent Hamiltonians derived in 
\cite{paperII}.

By way of illustration, in Figures 1 and 2 
we  plot $\langle x\rangle$ and $\langle p\rangle$
as functions of time for two of the systems 
$TM-\{1;b > -1;\omega=2;t_o=1\}$; namely, for 
$b=-0.5$ and $b=1.0$.  In Figure 1  we see that the envelope of the 
oscillation of 
$\langle x\rangle$ decreases with time while in Figure 2 we see 
that the envelope of the oscillation of 
$\langle p\rangle$ decreases with time. 
This ``exchange'' of 
maximum amplitude with time is a reflection of the coupling of 
$\langle x\rangle$ and $\langle p\rangle$ in the classical equations 
of motion [Eq. (\ref{tmem})]; i.e.,  the 
nonconservative nature of the forces.  This is emphasized by 
the phase-space plot in Figure 3 where, regardless of the initial position 
and momentum, the trajectory in phase space is an oscillatory motion 
about the phase-space origin with $\langle x\rangle$ 
becoming smaller as the 
amplitude of the oscillations in $\langle p\rangle$ become increasingly 
large.

As seen above, given that we have kept $\omega$ a constant, 
the frequency of the oscillation is higher the higher  
the value of $b$.  Contrariwise, if 
$b$ is fixed, the frequency of oscillation in $\langle x\rangle$ and 
$\langle p\rangle$ increases with increasing $\omega$.
 


\begin{figure}[p]
 \begin{center}
\noindent    
\psfig{figure=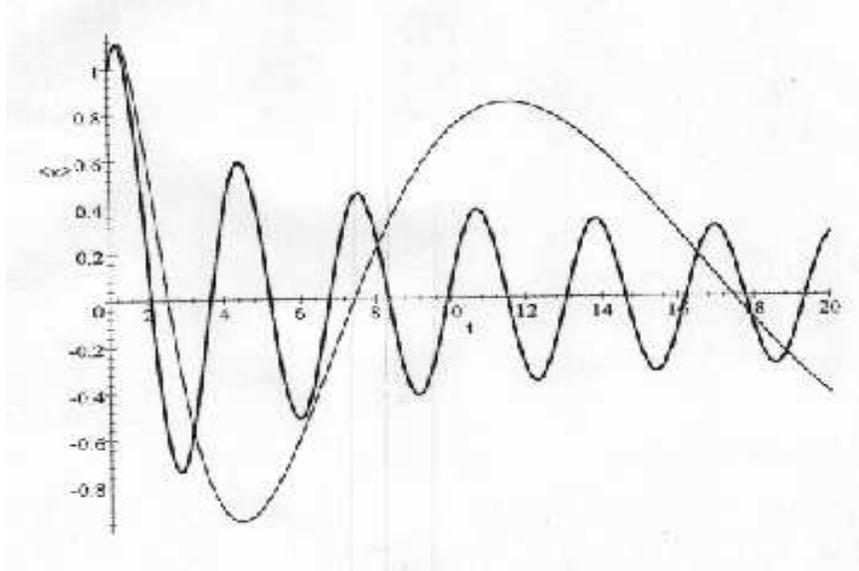,width=4.5in,height=3in}
  \caption{
 Plots of $\langle x\rangle$ versus time for the two systems: 
$TM-\{1;-0.5;\omega=2;t_o=1;x_o=1,p_o=1;\}$ (thin line) and 
$TM-\{1;1;\omega=2;t_o=1;x_o=1,p_o=1;\}$ (thick line).
 \label{fig:rodfig1}}
 \end{center}
\end{figure} 



\begin{figure}[p]
 \begin{center}
\noindent    
\psfig{figure=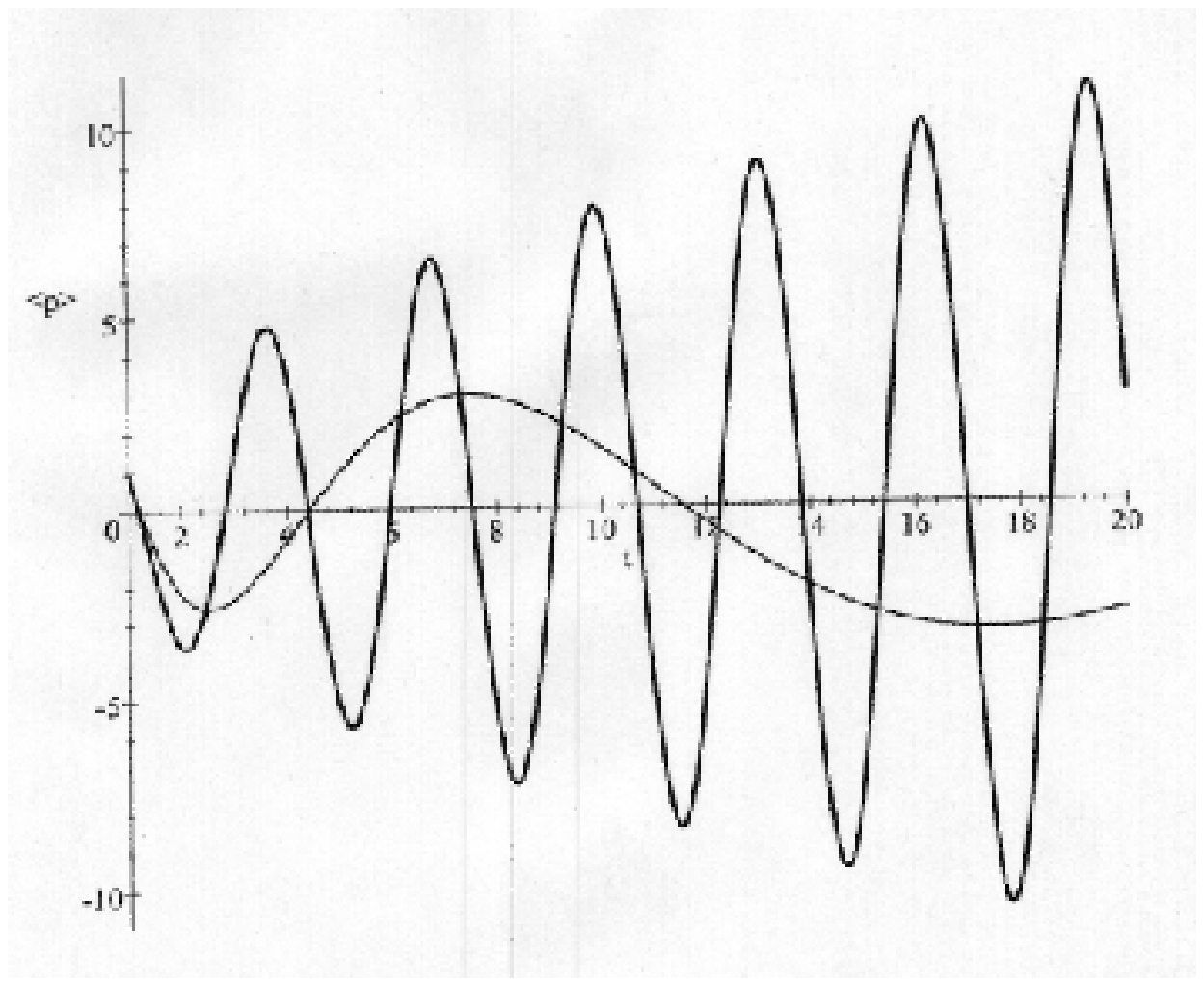,width=4.5in,height=3in}
  \caption{
Plots of $\langle p\rangle$ versus time for the two systems: 
$TM-\{1;-0.5;\omega=2;t_o=1;x_o=1,p_o=1;\}$ (thin line) and 
$TM-\{1;1;\omega=2;t_o=1;x_o=1,p_o=1;\}$ (thick line).
 \label{fig:rodfig2}}
 \end{center}
\end{figure} 



\begin{figure}[ht]
 \begin{center}
\noindent    
\psfig{figure=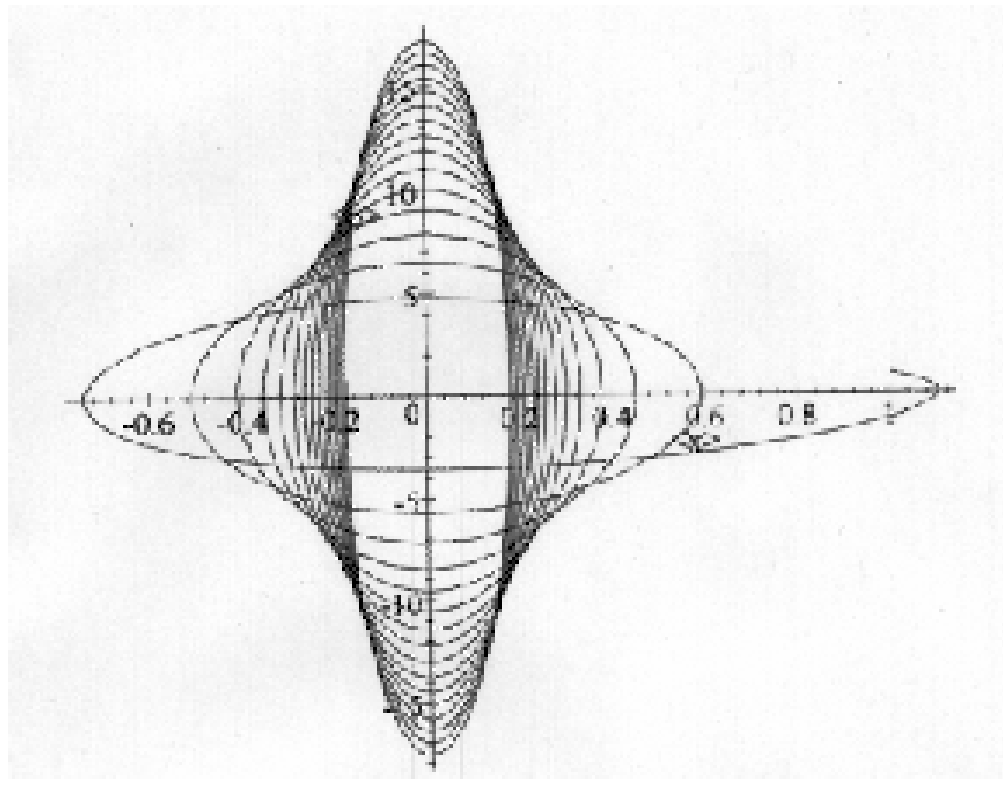,width=6in,height=4.5in}
  \caption{
A phase-space plot for the system 
$TM-\{1;1;\omega=2;t_o=1;x_o=1,p_o=1;\}$, for  time $1 \le t\le 50$.
 \label{fig:rodfig3}}
 \end{center}
\end{figure} 



\subsection{Uncertainties}

As with the expectation values in Section 6.1, 
uncertainties in position and momentum and the uncertainty products 
for all $TQ$, $TM$, and $TO$ systems, respectively,  
can be calculated from the general equations   
(106), (107), and (109) - (111) of \cite{paperII}, respectively  
using the functions from the 
appropriate tables in Appendix B.  

{\bf An illustration 
for $TM-\{1;\ne 1;\}$ systems:}  The uncertainties are    
\begin{eqnarray}
(\Delta x)^2 & = & \frac{\pi t_o}{4|b+1|}\left\{\left[\left(J_0^2(\sigma)-
Y_0^2(\sigma)\right)\cos{\theta}-J_0(\sigma)Y_0(\sigma)\sin{\theta}
\right]\sinh{2r}\right.\nonumber\\
  &   & ~~~~~~~~~~\left.+\left[J_0^2(\sigma)+Y_0^2(\sigma)\right]\cosh{2r}
\right\}\nonumber\\*[1mm]
(\Delta p)^2 & = & \frac{\pi|b+1|}{8t_o}\sigma\left\{\left[\left(
J_{-1}^2(\sigma)-Y_{-1}^2(\sigma)\right)\cos{\theta}-J_{-1}(\sigma)
Y_{-1}(\sigma)\sin{\theta}\right]\sinh{2r}\right.\nonumber\\
 &  & ~~~~~~~~~~\left.+\left[J_{-1}^2(\sigma)+
Y_{-1}^2(\sigma)\right]\cosh{2r}\right\}.\label{uxp}
\end{eqnarray}
This means the uncertainty product is
\begin{eqnarray}
(\Delta x)^2(\Delta p)^2 & = & \lfrac{1}{4}\left\{1+\lfrac{\pi^2\sigma^2}{4}
\left\{\left[J_{-1}(\sigma)J_0(\sigma)+Y_{-1}(\sigma)Y_0(\sigma)\right]
\cosh{2r}\right.\right.\nonumber\\
 &  & ~~~~~\left.\left.+\left[\left(J_0(\sigma)J_{-1}(\sigma)-Y_0(\sigma)Y_{-1}
(\sigma)\right)\cos{\theta}\right.\right.\right.\nonumber\\
 &  & ~~~~~\left.\left.\left.+\left(J_0(\sigma)Y_{-1}(\sigma)+Y_0(\sigma)
J_{-1}(\sigma)\right)\sin{\theta}\right]\sinh{2r}\right\}^2\right\}.
\label{upxp}
\end{eqnarray}   


\section{CONCLUSION}

For this type of Schr\"odinger system, the solutions obtained here 
apparently form a complete analysis for all real powers, $a$, 
$a\ne 0$, and $b$.  Further, the specific values of and/or ranges of 
$(a,b)$ for which specific solutions are valid was also determined.  
Thus, in toto, our results significantly extend the original work of 
Kim {\cite{spk1}}.  

The algebraic methods employed to obtain these solutions are 
based on the existence of isomorphic Lie symmetry algebras for $TO$-, $TM$-, 
and $TQ$-type Schr\"odinger equations.  
Lie symmetry analysis presents a powerful technique. 
With these symmetries we constructed a complete set of discrete states 
as well as coherent states and squeezed states.  

We emphasize again that the (time-dependent) discrete or number states 
are not eigenfunctions of the Hamiltonian. 
Yet further, the expectation values for $\langle x\rangle$ and 
$\langle p\rangle$ obtained in our analysis satisfy  Hamilton's 
equations of motion.


\section*{ACKNOWLEDGEMENTS}

MMN acknowledges the support of the United States Department of 
Energy.  DRT acknowledges
a grant from the Natural Sciences and Engineering Research Council 
of Canada.


\section*{APPENDIX A}
We give relationships involving Bessel and Hankel functions 
that are useful in performing calculations.  For details, we refer the 
reader to Lebedev {\cite{nnl1}}.  Below, $z$ is 
the independent variable and the prime indicates differentiation by $z$.
Wronskians are:
\begin{eqnarray}
W_z(J_{\mu},Y_{\mu})&=&J_{\mu}Y_{\mu}'-J_{\mu}'Y_{\mu}=\frac{2}{\pi z}.
\label{appa1}  \\
W_z(H_{\mu}^{(1)},H_{\mu}^{(2)})&=&H_{\mu}^{(1)}H_{\mu}^{(2)'}-
H_{\mu}^{(1)'}H_{\mu}^{(2)}=-\frac{4i}{\pi z}.
\label{appa4}
\end{eqnarray}
\noindent Let ${\cal F}_{\mu}(z)$ be a generic symbol for a Bessel 
function of the first of second kind, or a Hankel function.  Then, we 
have the following recursion relations:
\begin{eqnarray}
 & {\cal F}_{\mu-1}(z)+{\cal F}_{\mu+1}(z)=\frac{2\mu}{z}
{\cal F}_{\mu}(z), & \label{appa8}\\*[1mm]
 & 2{\cal F}_{\mu}'(z)={\cal F}_{\mu-1}(z)-{\cal F}_{\mu+1}(z), & 
\label{appa12}\\*[1mm]
 & {\cal F}_{\mu}'(z)={\cal F}_{\mu-1}(z)-\frac{\mu}{z}{\cal F}_{\mu}'(z).
 & \label{appa16}
\end{eqnarray}
\noindent The following are obtained by combining the 
Wronskians and recursion relations:
 \begin{eqnarray}
 & J_{\mu}(z)Y_{\mu-1}(z)-J_{\mu-1}(z)Y_{\mu}(z) = \frac{2}{\pi z}. & 
\label{appa20}\\*[1mm]
 & H_{\mu}^{(1)}(z)H_{\mu-1}^{(2)}(z)-H_{\mu-1}^{(1)}(z)H_{\mu}^{(2)} 
= -\frac{4i}{\pi z}. \label{appa24}
\end{eqnarray}


\section*{APPENDIX B}

\begin{center}


\begin{tabular}{|rcl|}
\multicolumn{3}{l}{Table B-1.  $\xi(t')$ and $\dot{\xi}(t')$ for
$TO$-systems.   
Variables and parameters are defined}\\
\multicolumn{3}{l}{as follows: 
$\sigma=\frac{2\omega t_o}{|b+1|}\exp\left[\frac{b+1}{2t_o}(t'-t_o')\right]$; 
$v=1+\frac{1-a}{t_o}(t'-t_o')$; $\tau=\frac{2\omega t_o}{|b-a+2|}v^{q/2}$;}\\
\multicolumn{3}{l}{$q=\frac{b-a+2}{1-a}$, and 
$\Delta^2=\left|1-\frac{4\omega^2 t_o^2}{(1-a)^2}\right|$. }\\*[1.5mm]
\hline\hline
 & & \\*[-3mm]
\multicolumn{1}{|c}{$TO$ System} & & \multicolumn{1}{c|}{$\xi(t')$} 
\\*[1mm]\hline
 & & \\*[-3mm]
$\scsz{\{1;(-1,\infty);\}}$ & \hspace{2cm} & $\scsz{\sqrt{
\lfrac{\pi t_o}{2|b+1|}}H^{(1)}_0\left(\sigma\right)}$ \\*[1.5mm]
$\scsz{\{1;(-\infty,-1);\}}$ & & $\scsz{\sqrt{\lfrac{\pi t_o}{2|b+1|}}
\bar{H}^{(1)}_0\left(\sigma\right)}$ \\*[2mm]
$\scsz{\{1,-1;\}}$ & & $\scsz{\sqrt{\lfrac{1}{2\omega}}\,
e^{i\omega(t'-t_o')}}$ \\*[2mm]
$\scsz{\{\ne 1;(a-2,\infty);\}}$ & & $\scsz{\sqrt{\frac{\pi t_o}{2|b-a+2|}}
\sqrt{v}H^{(1)}_{\frac{1}{q}}(\tau)}$ \\*[1.5mm]
$\scsz{\{\ne 1;(-\infty,a-2);\}}$ & & $\scsz{\sqrt{\lfrac{\pi t_o}{2|b-a+2|}}
\sqrt{v}\bar{H}^{(1)}_{\frac{1}{q}}(\tau)}$ \\*[2mm]
$\scsz{\{\ne 1;a-2;t_o<\frac{|1-a|}{2\omega};\pm;\}}$ & & $\scsz{
\sqrt{\lfrac{t_o}{2|1-a|\Delta}}\sqrt{v}\left(e^{-\frac{\Delta}{2}
\ln{v}}\pm ie^{\frac{\Delta}{2}\ln{v}}\right)}$ \\*[2mm]
$\scsz{\{\ne 1;a-2;t_o=\frac{|1-a|}{2\omega};\pm;\}}$ & & $\scsz{
\sqrt{\frac{t_o}{2|1-a|}}\sqrt{v}\left(1\pm i\ln{v}\right)}$ \\*[2mm]
$\scsz{\{\ne 1;a-2;t_o>\frac{|1-a|}{2\omega};\pm;\}}$ & & $\scsz{
\sqrt{\frac{t_o}{|1-a|\Delta}}\sqrt{v}\,e^{\pm i\frac{\Delta}{2}\ln{v}}}$ 
\\*[2mm]\hline
 & & \\*[-3mm]
\multicolumn{1}{|c}{$TO$ System} & & \multicolumn{1}{c|}{$\dot{\xi}(t')$} 
\\*[1mm]\hline
 & & \\*[-3mm]
$\scsz{\{1;(-1,\infty);\}}$ & & $\scsz{\lfrac{1}{2}\sqrt{
\lfrac{\pi|b+1|}{2t_o}}\sigma H^{(1)}_{-1}\left(\sigma\right)}$ \\*[2mm]
$\scsz{\{1;(-\infty,-1);\}}$ & & $\scsz{-\lfrac{1}{2}\sqrt{
\lfrac{\pi|b+1|}{2t_o}}\sigma\bar{H}^{(1)}_{-1}\left(\sigma\right)}$ \\*[2mm]
$\scsz{\{1;-1;\}}$ & & $\scsz{i\omega\sqrt{\lfrac{1}{2\omega}}\,
e^{i\omega(t'-t_o')}}$\\*[2mm]
$\scsz{\{\ne 1;(a-2,\infty);\}}$ & & $\scsz{\lfrac{1}{2}
\sqrt{\frac{\pi|b-a+2|}{2t_o}}\frac{1}{\sqrt{v}}\tau H^{(1)}_{\frac{1}{q}-1}
(\tau)}$\\*[2mm]
$\scsz{\{\ne 1;(-\infty,a-2);\}}$ & & $\scsz{-\lfrac{1}{2}\sqrt
{\frac{\pi|b-a+2|}{2t_o}}\frac{1}{\sqrt{v}}\tau\bar{H}^{(1)}_{\frac{1}{q}-1}
(\tau)}$\\*[2mm]
$\scsz{\{\ne 1;a-2;t_o<\frac{|1-a|}{2\omega};\pm;\}}$ & & $\scsz{\lfrac{1}{2}
\sqrt{\frac{|1-a|}{2t_o\Delta}}\frac{1}{\sqrt{v}}
\left[\pm(1-\Delta)e^{-\frac{\Delta}{2}\ln{v}}+i(1+\Delta)e^{\frac{\Delta}{2}
\ln{v}}\right]}$ \\*[1.5mm]
$\scsz{\{\ne 1;a-2;t_o=\frac{|1-a|}{2\omega};\pm;\}}$ & & $\scsz{
\sqrt{\frac{|1-a|}{2t_o}}\frac{1}{\sqrt{v}}\left[\pm\lfrac{1}{2}+
i\left(1+\lfrac{1}{2}\ln{v}\right)\right]}$ \\*[1.5mm]
$\scsz{\{\ne 1;a-2;t_o>\frac{|1-a|}{2\omega};\pm;\}}$ & & $\scsz{\lfrac{1}{2}
\sqrt{\lfrac{|1-a|}{t_o\Delta}}\frac{1}{\sqrt{v}}
(\pm 1+i\Delta)e^{\pm i\frac{\Delta}{2}\ln{v}}}$ \\*[4mm]\hline\hline
\end{tabular}



\begin{tabular}{|rcl|}
\multicolumn{3}{l}{Table B-2. $\phi_3(t')$, $\dot{\phi}_3(t')$, and 
$\ddot{\phi}_3(t')$ for $TO$-systems.  Variables and parameters}\\  
\multicolumn{3}{l}{are defined as follows 
$\sigma=\frac{2\omega t_o}{|b+1|}\exp\left[\frac{b+1}{2}(t'-t_o')\right]$;
$v=1+\frac{1-a}{t_o}(t'-t_o')$;}\\
\multicolumn{3}{l}{$\tau=\frac{2\omega t_o}{|b-a+2|}v^{q/2}$;  
$q=\frac{b-a+2}{1-a}$, and 
$\Delta^2=\left|1-\frac{4\omega^2t_o^2}{(1-a)^2}\right|$.}
\\*[1.5mm]\hline\hline
 & & \\*[-3mm]
\multicolumn{1}{|c}{$TO$ System} & & \multicolumn{1}{c|}{$\phi_3(t')$} 
\\*[1mm]\hline
 & & \\*[-3mm]
$\scsz{\{1;\ne -1;\}}$ & \hspace{2cm} & $\scsz{\lfrac{\pi t_o}{|b+1|}
H^{(1)}_0(\sigma)\bar{H}^{(1)}_0(\sigma)}$ \\*[2mm]
$\scsz{\{1;-1;\}}$ & & $\scsz{\lfrac{1}{\omega}}$\\*[2mm]
$\scsz{\{\ne 1;\ne a-2;\}}$ & & $\scsz{\frac{\pi t_o}{|b-a+2|}v 
H^{(1)}_{\frac{1}{q}}(\tau)\bar{H}^{(1)}_{\frac{1}{q}}(\tau)}$ \\*[2mm]
$\scsz{\{\ne 1;a-2;t_o<\frac{|1-a|}{2\omega};\pm\}}$ & & 
$\scsz{\frac{t_o}{|1-a|\Delta}v\left(e^{-\Delta\ln{v}}+e^{\Delta\ln{v}}
\right)}$ \\*[1.5mm]
$\scsz{\{\ne 1;a-2;t_o=\frac{|1-a|}{2\omega};\pm\}}$ & & $\scsz{
\frac{t_o}{|1-a|}v(1+\ln^2{v})}$ \\*[1.5mm]
$\scsz{\{\ne 1;a-2;t_o>\frac{|1-a|}{2\omega};\pm\}}$ & & $\scsz{
\frac{2t_o}{|1-a|\Delta}v}$ \\*[2mm]\hline
 & & \\*[-3mm]
\multicolumn{1}{|c}{$TO$ System} & & \multicolumn{1}{c|}{$\dot{\phi}_3(t')$} 
\\*[1mm]\hline
 & & \\*[-3mm]
$\left.\begin{array}{r}\scsz{\{1;(-1,\infty);\}}\\
\scsz{\{1;(-\infty,-1);\}}\end{array}\right\}$ & & $\scsz{\pm\frac{\pi}{2}
\sigma\left[H_{-1}^{(1)}(\sigma)\bar{H}_{0}^{(1)}(\sigma)+H_0^{(1)}(\sigma)
\bar{H}_{-1}^{(1)}(\sigma)\right]}$ \\*[2mm]
$\scsz{\{-1;-1;\}}$ & & $\scsz{0}$ \\*[2mm]
$\left.\begin{array}{r}\scsz{\{\ne 1;(a-2,\infty);\}}\\
\scsz{\{\ne 1;(-\infty,a-2);\}}\end{array}\right\}$ & & $\scsz{\pm
\frac{\pi}{2}\tau\left[H^{(1)}_{\frac{1}{q}-1}(\tau)\bar{H}^{(1)}_{
\frac{1}{q}}(\tau)+H^{(1)}_{\frac{1}{q}}(\tau)\bar{H}^{(1)}_{\frac{1}{q}-1}
(\tau)\right]}$ \\*[2mm]
$\scsz{\{\ne 1;a-2;t_o<\frac{|1-a|}{2\omega};\pm;\}}$ & & $\scsz{
\pm\lfrac{1}{\Delta}\left[(1-\Delta)e^{-\Delta\ln{s}}+(1+\Delta)
e^{\Delta\ln{s}}\right]}$ \\*[2mm]
$\scsz{\{\ne 1;a-2;t_o=\frac{|1-a|}{2\omega};\pm;\}}$ & & $\scsz{
\pm(1+\ln{s})^2}$ \\*[2mm]
$\scsz{\{\ne 1;a-2;t_o>\frac{|1-a|}{2\omega};\pm;\}}$ & & $\scsz{
\pm\lfrac{2}{\Delta}}$ \\*[2mm]\hline
 & & \\*[-3mm]
\multicolumn{1}{|c}{$TO$ System} & & \multicolumn{1}{c|}
{$\ddot{\phi}_3(t')$} \\*[1mm]\hline
 & & \\*[-3mm]
$\scsz{\{1;\ne -1;\}}$ & & $\scsz{\frac{\pi|b+|}{2t_o}\sigma^2\left[
H_{-1}^{(1)}(\sigma)\bar{H}_{-1}^{(1)}(\sigma)-H_0^{(1)}(\sigma)
\bar{H}_{0}^{(1)}(\sigma)\right]}$ \\*[2mm]
$\scsz{\{1;-1;\}}$ & & $\scsz{0}$\\*[2mm]
$\scsz{\{\ne 1;\ne a-2;\}}$ & & $\scsz{\frac{\pi|b-a+2|}{2t_o}\frac{1}{v}
\tau^2\left[H^{(1)}_{\frac{1}{q}-1}(\tau)\bar{H}^{(1)}_{\frac{1}{q}-1}(\tau)
-H^{(1)}_{\frac{1}{q}}(\tau)\bar{H}^{(1)}_{\frac{1}{q}}(\tau)\right]}$ \\*[2mm]
$\scsz{\{\ne 1;a-2;t_o<\frac{|1-a|}{2\omega};\pm;\}}$ & & $\scsz{
\frac{|1-a|}{t_o}\frac{1}{v}\left[-(1-\Delta)e^{-\Delta\ln{v}}
+(1+\Delta)e^{\Delta\ln{v}}\right]}$ \\*[2mm]
$\scsz{\{\ne 1;a-2;t_o=\frac{|1-a|}{2\omega};\pm;\}}$ & & $\scsz{
\frac{2|1-a|}{t_o}\frac{1}{v}\left(1+\ln{v}\right)}$ \\*[1.5mm]
$\scsz{\{\ne 1;a-2;t_o>\frac{|1-a|}{2\omega};\pm;\}}$ & & $\scsz{0}$ \\
 & & \\\hline\hline
\end{tabular}


 
\begin{tabular}{|rcl|}
\multicolumn{3}{l}{Table B-3.  $\hat{\xi}(t)$ and $\hat{\dot{\xi}}(t)$ 
for $TM$-systems.  Variables and parameters are defined }\\
\multicolumn{3}{l}{as follows: $\tau=\frac{2\omega t_o}{|b-a+2|}
\left(\frac{t}{t_o}\right)^{\frac{b-a+2}{2}}$; 
$\chi=\frac{1}{2}(1-a)\ln{\frac{t}{t_o}}$; $q=\frac{b-a+2}{1-a}$, }\\  
\multicolumn{3}{l}{and $\Delta^2=\left|1-\frac{4\omega^2t_o^2}{(1-a)^2}
\right|$.}\\*[2mm]\hline\hline
 & & \\*[-3mm]
\multicolumn{1}{|c}{$TM$ System} & & \multicolumn{1}{c|}{$\hat{\xi}(t)$} 
\\*[1mm]\hline
 & & \\*[-3mm]
$\scsz{\{\ne 0;(a-2,\infty);\}}$ & \hspace{2cm} & $\scsz{\sqrt{\frac{\pi t_o}
{2|b-a+2|}}\left(\frac{t}{t_o}\right)^{\frac{1-a}{2}}H^{(1)}_{\frac{1}{q}}
(\tau)}$ \\*[2mm]
$\scsz{\{\ne 0;(-\infty,a-2);\}}$ & & $\scsz{\sqrt{\lfrac{\pi t_o}
{2|b-a+2|}}\left(\frac{t}{t_o}\right)^{\frac{1-a}{2}}
\bar{H}^{(1)}_{\frac{1}{q}}(\tau)}$  \\*[2mm]
$\scsz{\{1;-1;\}}$ & & $\scsz{\sqrt{\frac{1}{2\omega}}\,
e^{i\omega t_o\ln{\left(\frac{t}{t_o}\right)}}}$ \\*[2mm]
$\scsz{\{\ne 1;a-2;t_o<\frac{|1-a|}{2\omega};\pm;\}}$ & & $\scsz{
\sqrt{\frac{t_o}{2|1-a|\Delta}}\left(\frac{t}{t_o}
\right)^{\frac{1-a}{2}}\left(e^{-\Delta\chi}
\pm ie^{\Delta\chi}\right)}$ \\*[2mm]
$\scsz{\{\ne 1;a-2;t_o=\frac{|1-a|}{2\omega};\pm;\}}$ & & $\scsz{
\sqrt{\frac{t_o}{2|1-a|}}\left(\frac{t}{t_o}\right)^{\frac{1-a}{2}}
\left(1\pm 2i\chi\right)}$  \\*[2mm]
$\scsz{\{\ne 1;a-2;t_o>\frac{|1-a|}{2\omega};\pm;\}}$ & & $\scsz{
\sqrt{\frac{t_o}{|1-a|\Delta}}\left(\frac{t}{t_o}
\right)^{\frac{1-a}{2}}\,e^{\pm i\Delta\chi}}$  \\*[2mm]\hline
 & & \\*[-3mm]
\multicolumn{1}{|c}{$TM$ Systems} & & \multicolumn{1}{c|}
{$\hat{\dot{\xi}}(t)$} \\*[1mm]\hline
 & & \\*[-3mm]
$\scsz{\{\ne 0;(a-2,\infty);\}}$ & & $\scsz{\frac{1}{2}\sqrt{
\frac{\pi|b-a+2|}{2t_o}}\left(\frac{t}{t_o}\right)^{\frac{a-1}{2}}\tau 
H^{(1)}_{\frac{1}{q}-1}(\tau)}$ \\*[2mm]
$\scsz{\{\ne 0;(-\infty,a-2);\}}$ & & $\scsz{-\frac{1}{2}\sqrt{
\frac{\pi|b-a+2|}{2t_o}}\left(\frac{t}{t_o}
\right)^{\frac{a-1}{2}}\tau\bar{H}^{(1)}_{\frac{1}{q}-1}(\tau)}$
\\*[2mm]
$\scsz{\{1;-1;\}}$ & & $\scsz{i\omega\sqrt{\lfrac{1}{2\omega}}\,
e^{i\omega t_o\ln{\frac{t}{t_o}}}}$ \\*[2mm]
$\scsz{\{\ne 1;a-2;t_o<\frac{|1-a|}{2\omega};\pm;\}}$ & & $\scsz{\frac{1}{2}
\sqrt{\frac{|1-a|}{2t_o\Delta}}\left(\frac{t}{t_o}\right)^{\frac{a-1}{2}}
\left[\pm(1-\Delta)e^{-\Delta\chi}+i(1+\Delta)e^{\Delta\chi}\right]}$ 
\\*[2mm]
$\scsz{\{\ne 1;a-2;t_o=\frac{|1-a|}{2\omega};\pm;\}}$ & & $\scsz{
\sqrt{\frac{|1-a|}{2t_o}}\left(\frac{t}{t_o}\right)^{\frac{a-1}{2}}
\left[\pm\lfrac{1}{2}+i\left(1+\chi\right)\right]}$  \\*[1.5mm]
$\scsz{\{\ne 1;a-2;t_o>\frac{|1-a|}{2\omega};\pm;\}}$ & & $\scsz{
\frac{1}{2}\sqrt{\frac{|1-a|}{t_o\Delta}}\left(\frac{t}{t_o}\right)
^{\frac{a-1}{2}}(\pm 1+i\Delta)e^{\pm i\Delta\chi}}$  \\
 & & \\\hline\hline
\end{tabular}



\begin{tabular}{|rcl|}
\multicolumn{3}{l}{Table B-4. $\hat{\phi}_3(t)$, $\hat{\dot{\phi}}_3(t)$, and 
$\hat{\ddot{\phi}}_3(t)$ for $TM$-systems.  Variables and parameters are }\\
\multicolumn{3}{l}{defined as follows: $\tau=\frac{2\omega t_o}{|b-a+2|}
\left(\frac{t}{t_o}\right)^{\frac{b-a+2}{2}}$; 
$\chi=\frac{1}{2}(1-a)\ln{\frac{t}{t_o}}$; $q=\frac{b-a+2}{1-a}$,}\\
\multicolumn{3}{l}{and 
$\Delta^2=\left|1-\frac{4\omega^2t_o^2}{(1-a)^2}\right|$.}\\*[1.5mm]
\hline\hline
 & & \\*[-3mm]
\multicolumn{1}{|c}{$TM$ System} & & \multicolumn{1}{c|}{$\hat{\phi}_3(t)$} 
\\*[1mm]\hline
 & & \\*[-3mm]
$\scsz{\{\ne 0;\ne a-2;\}}$ & \hspace{2cm} & $\scsz{\frac{\pi t_o}
{|b-a+2|}\left(\frac{t}{t_o}\right)^{1-a} 
H^{(1)}_{\frac{1}{q}}(\tau)\bar{H}^{(1)}_{\frac{1}{q}}(\tau)}$  \\*[2mm]
$\scsz{\{1;-1\}}$ & & $\scsz{\frac{1}{\omega}}$ \\*[2mm]
$\scsz{\{\ne 1;a-2;t_o<\frac{|1-a|}{\omega};\pm;\}}$ & & $\scsz{
\frac{t_o}{|1-a|\Delta}\left(\frac{t}{t_o}\right)^{1-a}
\left(e^{-2\Delta\chi}+e^{2\Delta\chi}\right)}$  \\*[2mm]
$\scsz{\{\ne 1;a-2;t_o=\frac{|1-a|}{\omega};\pm;\}}$ & & $\scsz{
\frac{t_o}{|1-a|}\left(\frac{t}{t_o}\right)^{1-a}(1+4\chi^2)}$  \\*[2mm]
$\scsz{\{\ne 1;a-2;t_o>\frac{|1-a|}{\omega};\pm;\}}$ & & $\scsz{
\frac{2t_o}{|1-a|\Delta}\left(\frac{t}{t_o}\right)^{1-a}}$  
\\*[2mm]\hline
 & & \\*[-3mm]
\multicolumn{1}{|c}{$TM$ System} & & \multicolumn{1}{c|}{$
\hat{\dot{\phi}}_3(t)$} \\*[1mm]\hline
 & & \\*[-3mm]
$\left.\begin{array}{r}\scsz{\{\ne 0;(a-2,\infty);\}}^+\\
\scsz{\{\ne 0;(-\infty,a-2);\}^-}\end{array}\right\}$ & & $\scsz{\pm
\frac{\pi}{2}\tau\left[H^{(1)}_{\frac{1}{q}-1}(\tau)\bar{H}^{(1)}_
{\frac{1}{q}}(\tau)+H^{(1)}_{\frac{1}{q}}(\tau)\bar{H}^{(1)}_{\frac{1}{q}-1}
(\tau)\right]}$ \\*[2mm] 
$\scsz{\{1;-1;\}}$ & & $\scsz{0}$ \\*[2mm]
$\scsz{\{\ne 1;a-2;t_o<\frac{|1-a|}{2\omega};\pm;\}}$ & & $\scsz{\pm\
\frac{1}{\Delta}\left[(1-\Delta)e^{-2\Delta\chi}+(1+\Delta)e^{2\Delta\chi}
\right]}$  \\*[2mm]
$\scsz{\{\ne 1;a-2;t_o=\frac{|1-a|}{2\omega};\pm;\}}$ & & 
$\scsz{\pm(1+2\chi)^2}$ \\*[2mm]
$\scsz{\{\ne 1;a-2;t_o>\frac{|1-a|}{2\omega};\pm;\}}$ & & $\scsz{\pm
\frac{2}{\Delta}}$ \\*[2mm]\hline
 & & \\*[-3mm]
\multicolumn{1}{|c}{$TM$ System} & & \multicolumn{1}{c|}{$
\hat{\ddot{\phi}}_3(t)$} \\*[1mm]\hline
 & & \\*[-3mm]
$\scsz{\{\ne 0;\ne a-2;\}}$ & & $\scsz{
\frac{\pi|b-a+2|}{2t_o}\left(\frac{t}{t_o}\right)^{a-1}\tau^2\left[
H^{(1)}_{\frac{1}{q}-1}(\tau)\bar{H}^{(1)}_{\frac{1}{q}-1}(\tau)-
H^{(1)}_{\frac{1}{q}}(\tau)\bar{H}^{(1)}_{\frac{1}{q}}(\tau)\right]}$  
\\*[2mm]
$\scsz{\{1;-1;\}}$ & & $\scsz{0}$\\*[2mm]
$\scsz{\{\ne 1;a-2;t_o<\frac{|1-a|}{2\omega};\pm;\}}$ & & $\scsz{
\frac{|1-a|}{t_o}\left(\frac{t}{t_o}\right)^{a-1}\left[
-(1-\Delta)e^{-2\Delta\chi}+(1+\Delta)e^{2\Delta\chi}\right]}$  
\\*[2mm]
$\scsz{\{\ne 1;a-2;t_o=\frac{|1-a|}{2\omega};\pm;\}}$ & & $\scsz{
\frac{2|1-a|}{t_o}\left(\frac{t}{t_o}\right)^{a-1}\left(1+
2\chi\right)}$  \\*[2mm]
$\scsz{\{\ne 1;a-2;t_o=\frac{|1-a|}{2\omega};\pm;\}}$ & & $\scsz{0}$ \\
 & & \\\hline\hline
\end{tabular}



\begin{tabular}{|rcl|}
\multicolumn{3}{l}{Table B-5.  $\Xi_P(t)$ and $\Xi_X(t)$ 
for $TQ$-systems.  Variables and parameters are }\\
\multicolumn{3}{l}{defined as follows: $\tau=\frac{2\omega t_o}{|b-a+2|}
\left(\frac{t}{t_o}\right)^{\frac{b-a+2}{2}}$; 
$\chi=\frac{1}{2}(1-a)\ln{\frac{t}{t_o}}$; $q=\frac{b-a+2}{1-a}$, }\\
\multicolumn{3}{l}{and 
$\Delta^2=\left|1-\frac{4\omega^2t_o^2}{(1-a)^2}\right|$.}\\*[1.5mm]
\hline\hline
 & & \\*[-3mm]
\multicolumn{1}{|c}{$TQ$ System} & & \multicolumn{1}{c|}{$\Xi_P(t)$} 
\\*[1mm]\hline
 & & \\*[-3mm]
$\scsz{\{\ne 0;(a-2,\infty;\}}$ & \hspace{2cm} & $\scsz{\sqrt{
\frac{\pi|t_o}{2|b-a+2|}}
\left(\frac{t}{t_o}\right)^{\frac{1}{2}}H^{(1)}_{1/q}(\tau)}$  \\*[2mm]
$\scsz{\{\ne 0;(-\infty,a-2);\}}$ & & $\scsz{\sqrt{\frac{\pi t_o}{2|b-a+2|}}
\left(\frac{t}{t_o}\right)^{\frac{1}{2}}\bar{H}^{(1)}_{1/q}(\tau)}$ \\*[2mm]
$\scsz{\{1;-1;\}}$ & & $\scsz{\sqrt{\lfrac{1}{2\omega}}\left(\frac{t}{t_o}
\right)^{\frac{1}{2}}\,e^{i\omega t_o\ln{\left(\frac{t}{t_o}\right)}}}$ 
\\*[2mm]
$\scsz{\{\ne 0,1;a-2;t_o<\frac{|1-a|}{2\omega};\pm;\}}$ & & $\scsz{
\sqrt{\frac{t_o}{2|1-a|\Delta}}\left(\frac{t}{t_o}\right)^{\frac{1}{2}}
\left(e^{-\Delta\chi}\pm ie^{\Delta\chi}\right)}$ \\*[2mm]
$\scsz{\{\ne 0,1;a-2;t_o=\frac{|1-a|}{2\omega};\pm;\}}$ & & $\scsz{
\sqrt{\frac{t_o}{2|1-a|}}\left(\frac{t}{t_o}\right)^{\frac{1}{2}}
\left(1\pm 2i\chi\right)}$ \\*[2mm]
$\scsz{\{\ne 0,1;a-2;t_o>\frac{|1-a|}{2\omega};\pm;\}}$ & & $\scsz{
\sqrt{\frac{t_o}{|1-a|\Delta}}\left(\frac{t}{t_o}\right)^{\frac{1}{2}}
\,e^{\pm i\Delta\chi}}$ \\*[2mm]\hline
 & & \\*[-3mm]
\multicolumn{1}{|c}{$TQ$ System} & & \multicolumn{1}{c|}{$\Xi_X(t)$} 
\\*[1mm]\hline
 & & \\*[-3mm]
$\scsz{\{\ne 0;(a-2,\infty);\}}$ & & $\scsz{\frac{1}{2}\sqrt{
\frac{\pi|b-a+2|}{2t_o}}\left(\frac{t}{t_o}\right)^{-\frac{1}{2}}\tau 
H^{(1)}_{\frac{1}{q}-1}(\tau)}$ \\*[2mm]
$\scsz{\{\ne 0;(-\infty,a-2);\}}$ & & $\scsz{-\frac{1}{2}\sqrt{\frac{\pi|b-a+2|}{2t_o}}\left(\frac{t}{t_o}
\right)^{-\frac{1}{2}}\tau\bar{H}^{(1)}_{\frac{1}{q}-1}(\tau)}$ \\*[2mm]
$\scsz{\{1;-1;\}}$ & & $\scsz{i\omega\sqrt{\frac{1}{2\omega}}\left(
\frac{t}{t_o}\right)^{-\frac{1}{2}}\,e^{i\omega t_o\ln{\frac{t}{t_o}}}}$ 
\\*[2mm]
$\scsz{\{\ne 0,1;a-2;t_o<\frac{|1-a|}{2\omega};\pm;\}}$ & & $\scsz{
\frac{1}{2}\sqrt{\frac{|1-a|}{2t_o\Delta}}\left(\frac{t}{t_o}
\right)^{-\frac{1}{2}}\left[\pm(1-\Delta)e^{-\Delta\chi}
+i(1+\Delta)e^{\Delta\chi}\right]}$ \\*[2mm]
$\scsz{\{\ne 0,1;a-2;t_o=\frac{|1-a|}{2\omega};\pm;\}}$ & & $\scsz{
\sqrt{\frac{|1-a|}{2t_o}}\left(\frac{t}{t_o}\right)^{-\frac{1}{2}}
\left[\pm\lfrac{1}{2}+i\left(1+\chi\right)\right]}$ \\*[2mm]
$\scsz{\{\ne 0,1;a-2;t_o>\frac{|1-a|}{2\omega};\pm;\}}$ & & $\scsz{
\frac{1}{2}\sqrt{\frac{|1-a|}{t_o\Delta}}\left(\frac{t}{t_o}\right)
^{-\frac{1}{2}}(\pm 1+i\Delta)e^{\pm i\Delta\chi}}$ \\*[2mm]
 & & \\\hline\hline
\end{tabular}



\begin{tabular}{|rcl|}
\multicolumn{3}{l}{Table B-6. $C_{3,T}(t)$, $C_{3,D}(t)$, and 
$C_{3,X^2}(t)$ for $TQ$-systems.  Variables and }\\  
\multicolumn{3}{l}{parameters are defined as follows: 
$\tau=\frac{2\omega t_o}{|q|}\left(\frac{t}{t_o}\right)^{\frac{b-a+2}{2}}$; 
$\chi=\frac{1}{2}(1-a)\ln{\frac{t}{t_o}}$; }\\
\multicolumn{3}{l}{$q=\frac{b-a+2}{1-a}$, and 
$\Delta^2=\left|1-\frac{4\omega^2t_o^2}{(1-a)^2}\right|$.}
\\*[1.5mm]\hline\hline
 & & \\*[-3mm]
\multicolumn{1}{|c}{$TQ$ System} & & \multicolumn{1}{c|}{$C_{3,T}(t)$} 
\\*[1mm]\hline
 & & \\*[-3mm]
$\scsz{\{\ne 0;\ne a-2;\}}$ & & $\scsz{\lfrac{\pi t_o}{|b-a+2|}\left(
\frac{t}{t_o}\right) H^{(1)}_{\frac{1}{q}}(\tau)\bar{H}^{(1)}_{\frac{1}{q}}
(\tau)}$ \\*[2mm]
$\scsz{\{1;-1;\}}$ & & $\scsz{\lfrac{1}{\omega}\left(\frac{t}{t_o}\right)}$ 
\\*[2mm]
$\scsz{\{\ne 0,1;a-2;t_o<\frac{|1-a|}{2\omega};\pm\}}$ & & $\scsz{
\frac{t_o}{|1-a|\Delta}\left(\frac{t}{t_o}\right)
\left(e^{-2\Delta\chi}+e^{2\Delta\chi}\right)}$ \\*[2mm]
$\scsz{\{\ne 0,1;a-2;t_o=\frac{|1-a|}{2\omega};\pm\}}$ & & $\scsz{
\lfrac{|\beta|}{\omega}\left(\frac{t}{t_o}\right)(1+4\chi^2)}$ 
\\*[2mm]
$\scsz{\{\ne 0,1;a-2;t_o>\frac{|1-a|}{2\omega};\pm\}}$ & & $\scsz{
\lfrac{2|\beta|}{\omega\Delta}\left(\frac{t}{t_o}\right)}$ 
\\*[2mm]\hline
 & & \\*[-3mm]
\multicolumn{1}{|c}{$TQ$-System} & & \multicolumn{1}{c|}{$C_{3,D}(t)$} 
\\*[1mm]\hline
 & & \\*[-3mm]
$\left.\begin{array}{r}\scsz{\{\ne 0;(a-2,\infty);\}^+}\\
\scsz{\{\ne 0;(-\infty,a-2);\}^-}\end{array}\right\}$ & & $\scsz{
\frac{a\pi t_o}{2|b-a+2|}\left(\frac{1}{t_o}\right)H_{\frac{1}{q}}^{(1)}
(\tau)\bar{H}_{\frac{1}{q}}^{(1)}(\tau)\pm\lfrac{\pi}{4}\tau
\left[H^{(1)}_{\frac{1}{q}-1}(\tau)\bar{H}^{(1)}_{\frac{1}{q}}(\tau)+
H^{(1)}_{\frac{1}{q}}(\tau)\bar{H}^{(1)}_{\frac{1}{q}-1}(\tau)\right]}$  
\\*[2mm]
$\scsz{\{1;-1;\}}$ & & $\scsz{\frac{1}{2\omega t_o}}$ \\*[2mm]
$\scsz{\{\ne0,1;a-2;t_o<\frac{|1-a|}{2\omega};\pm;\}}$ & & $\scsz{
\frac{at_o}{2|1-a|\Delta}\left(\frac{1}{t_o}\right)\left(
e^{-2\Delta\chi}+e^{2\Delta\chi}\right)\pm\lfrac{1}{2\Delta}\left[
(1-\Delta)e^{-2\Delta\chi}+(1+\Delta)e^{2\Delta\chi}\right]}$  
\\*[2mm]
$\scsz{\{\ne0,1;a-2;t_o=\frac{|1-a|}{2\omega};\pm;\}}$ & & $\scsz{
\frac{at_o}{2|1-a|}\left(\lfrac{1}{t_o}\right)(1+4\chi^2)\pm 
\left[\frac{1}{2}+2\chi(1+\chi)\right]}$ \\*[2mm]
$\scsz{\{\ne0,1;a-2;t_o>\frac{|1-a|}{2\omega};\pm;\}}$ & & $\scsz{
\frac{at_o}{|1-a|\Delta}\left(\frac{1}{t_o}\right)\pm\frac{1}{\Delta}}$ 
\\*[2mm]\hline
 & & \\*[-3mm]
\multicolumn{1}{|c}{$TQ$ System} & & \multicolumn{1}{c|}{$C_{3,X^2}(t)$} 
\\*[1mm]\hline
 & & \\*[-3mm]
$\scsz{\{\ne 0;\ne a-2;\}}$ & & $\scsz{\frac{\pi|b-a+2|}{8t_o}\left(
\frac{t}{t_o}\right)^{-1}\tau^2\left[H^{(1)}_{\frac{1}{q}-1}(\tau)
\bar{H}^{(1)}_{\frac{1}{q}-1}(\tau)-H^{(1)}_{\frac{1}{q}}(\tau)
\bar{H}^{(1)}_{\frac{1}{q}}(\tau)\right]}$ \\*[2mm] 
$\scsz{\{1;-1;\}}$ & & $\scsz{0}$ \\*[2mm]
$\scsz{\{\ne 0,1;a-2;t_o<\frac{|1-a|}{2\omega};\pm;\}}$ & & $\scsz{
-\frac{|1-a|}{4t_o}\left(\frac{t}{t_o}\right)^{-1}\left[
-(1-\Delta)e^{-2\Delta\chi}+(1+\Delta)e^{2\Delta\chi}\right]}$  
\\*[2mm]
$\scsz{\{\ne 0,1;a-2;t_o=\frac{|1-a|}{2\omega};\pm;\}}$ & & $\scsz{
-\frac{|1-a|}{2t_o}\left(\frac{t}{t_o}\right)^{-1}\left(1+
2\chi\right)}$ \\*[2mm]
$\scsz{\{\ne 0,1;a-2;t_o>\frac{|1-a|}{2\omega};\pm;\}}$ & & $\scsz{0}$ \\
 & & \\\hline\hline
\end{tabular}

\end{center} 



\end{document}